\PassOptionsToPackage{table}{xcolor}
\documentclass[conference]{IEEEtran}

\usepackage{amsmath,amsfonts}

\usepackage{url}

\usepackage{tikz}
\usepackage{amsmath}
\usepackage{booktabs} %

\usepackage{xspace}
\usepackage[disable]{todonotes}
\usepackage{tcolorbox}
\usepackage{enumitem}

\usepackage{numprint}
\npthousandsep{\,}

\usepackage[]{hyperref}         %
\hypersetup{                    %
  colorlinks,
  linkcolor={green!80!black},
  citecolor={red!70!black},
  urlcolor={blue!70!black}
}

\usepackage{cleveref}
\usepackage{adjustbox}
\usepackage{algpseudocode}

\algtext*{EndWhile}
\algtext*{EndIf}
\algtext*{EndFor}
\algtext*{EndProcedure}
\algtext*{EndFunction}

\usepackage{algorithm}

\usepackage{pgfplots}

\usepackage{color}
\definecolor{lightgray}{rgb}{.9,.9,.9}
\definecolor{darkgray}{rgb}{.4,.4,.4}
\definecolor{purple}{rgb}{0.65, 0.12, 0.82}

\pgfplotscreateplotcyclelist{mycolorlist}{%
	red!40!white,            style={fill=red!20!white},   mark=none\\%
	brown!40!white,          style={fill=brown!20!white}, mark=none\\%
	blue!40!white,           style={fill=blue!20!white},  mark=none\\%
	black,mark=star\\%
	blue,every mark/.append style={fill=blue!80!black},mark=diamond*\\%
	red,densely dashed,every mark/.append style={solid,fill=red!80!black},mark=*\\%
	brown!60!black,densely dashed,every mark/.append style={
		solid,fill=brown!80!black},mark=square*\\%
	black,densely dashed,every mark/.append style={solid,fill=gray},mark=otimes*\\%
	blue,densely dashed,mark=star,every mark/.append style=solid\\%
	red,densely dashed,every mark/.append style={solid,fill=red!80!black},mark=diamond*\\%
}

\usepackage{listings}
\usepackage{xcolor}

\definecolor{dkgreen}{rgb}{0,.6,0}
\definecolor{dkblue}{rgb}{0,0,.6}
\definecolor{dkyellow}{cmyk}{0,0,.8,.3}
\definecolor{lightgray}{rgb}{0.95, 0.95, 0.95}

\usepackage{fixltx2e}

\usepackage{multirow}

\def\SPSB#1#2{\rlap{\textsuperscript{#1}}\SB{#2}}

\def\SB#1{\textsubscript{#1}}

\newcommand{\shortname}{SpiderSapien\xspace}
\newcommand{\ablatedLLM}{{\small SS\SPSB{-LLM}{+ELM}}\xspace}
\newcommand{\ablatedELM}{{\small SS\SPSB{-ELM}{+LLM}}\xspace}

\newcommand{\arachni}{Arachni\xspace}
\newcommand{\blackostrich}{Black Ostrich\xspace}
\newcommand{\wapiti}{Wapiti\xspace}
\newcommand{\zap}{ZAP\xspace}
\newcommand{\totalScanners}{6\xspace}

\newcommand{\jak}{j\"Ak\xspace}
\newcommand{\bw}{Black Widow\xspace}
\newcommand{\evocrawl}{EvoCrawl\xspace}
\newcommand{\yurascanner}{Yura\-Scanner\xspace} %
\newcommand{\yura}{\yurascanner} %

\newcommand{\technique}{immersive interaction\xspace}

\newcommand{\totalApplications}{9\xspace}
\newcommand{\xssApplications}{7\xspace} %

\newcommand{\dokuwiki}{DokuWiki\xspace}
\newcommand{\tinyfilemanager}{TinyFileManager\xspace}
\newcommand{\kanboard}{Kanboard\xspace}
\newcommand{\wordpress}{WordPress\xspace}
\newcommand{\oscommerce}{osCommerce\xspace}
\newcommand{\hotcrp}{HotCRP\xspace}
\newcommand{\leantime}{Leantime\xspace}
\newcommand{\piwigo}{Piwigo\xspace}
\newcommand{\nextcloud}{Nextcloud\xspace}

\newcommand{\totalXSS}{32\xspace}

\newcommand{\othersXSS}{29\xspace}
\newcommand{\othersXSSApplications}{3\xspace}

\newcommand{\exploitableXSS}{26\xspace}

\newcommand{\ablatedLLMTotalXSS}{28\xspace}
\newcommand{\ablatedLLMXSSApplications}{4\xspace}
\newcommand{\ablatedLLMOverlapXSS}{13\xspace}
\newcommand{\ablatedELMTotalXSS}{27\xspace}
\newcommand{\ablatedELMXSSApplications}{3\xspace}
\newcommand{\ablatedELMOverlapXSS}{15\xspace}
\newcommand{\yuraRepeatXSS}{11\xspace}
\newcommand{\yuraRepeatOverlapYuraXSS}{8\xspace}

\newcommand{\ourRepeatOscommerceXSS}{17\xspace}
\newcommand{\ourOscommerceXSS}{19\xspace}
\newcommand{\ourOscommerceOverlapXSS}{12\xspace}
\newcommand{\ourRepeatXSS}{31\xspace}
\newcommand{\ourRepeatXSSApplications}{6\xspace}

\newcommand{\clientCoveragePercent}{65\%\xspace}

\newcommand{\clientCoverageImprovedkanboard}{5\%\xspace}

\newcommand{\clientCoverageImprovedpiwigo}{1020\%\xspace}

\newcommand{\clientCoverageImprovedAvg}{287\%\xspace}

\newcommand{\improvedCoverageTwoONdokuwikiVSblackostrich}{19.69\%\xspace}

\newcommand{\improvedCoverageTwoONdokuwikiVSarachni}{154.06\%\xspace}

\newcommand{\improvedCoverageTwoONdokuwikiVSwapiti}{12.67\%\xspace}

\newcommand{\improvedCoverageTwoONdokuwikiVSevocrawl}{14.41\%\xspace}

\newcommand{\improvedCoverageTwoONdokuwikiVSzap}{3.86\%\xspace}

\newcommand{\improvedCoverageTwoONdokuwikiVSyura}{16.44\%\xspace}
\newcommand{\improvedCoverageVsAllOndokuwiki}{-12.24\%\xspace}

\newcommand{\improvedCoverageTwoONhotcrpVSblackostrich}{83.81\%\xspace}

\newcommand{\improvedCoverageTwoONhotcrpVSarachni}{745.81\%\xspace}

\newcommand{\improvedCoverageTwoONhotcrpVSwapiti}{746.99\%\xspace}

\newcommand{\improvedCoverageTwoONhotcrpVSevocrawl}{72.09\%\xspace}

\newcommand{\improvedCoverageTwoONhotcrpVSzap}{406.66\%\xspace}

\newcommand{\improvedCoverageTwoONhotcrpVSyura}{179.69\%\xspace}
\newcommand{\improvedCoverageVsAllOnhotcrp}{44.63\%\xspace}

\newcommand{\improvedCoverageTwoONtinyfilemanagerVSblackostrich}{40.61\%\xspace}

\newcommand{\improvedCoverageTwoONtinyfilemanagerVSarachni}{46.85\%\xspace}

\newcommand{\improvedCoverageTwoONtinyfilemanagerVSwapiti}{49.63\%\xspace}

\newcommand{\improvedCoverageTwoONtinyfilemanagerVSevocrawl}{7.40\%\xspace}

\newcommand{\improvedCoverageTwoONtinyfilemanagerVSzap}{-0.97\%\xspace}

\newcommand{\improvedCoverageTwoONtinyfilemanagerVSyura}{48.54\%\xspace}
\newcommand{\improvedCoverageVsAllOntinyfilemanager}{-5.64\%\xspace}

\newcommand{\improvedCoverageTwoONkanboardVSblackostrich}{85.04\%\xspace}

\newcommand{\improvedCoverageTwoONkanboardVSarachni}{153.47\%\xspace}

\newcommand{\improvedCoverageTwoONkanboardVSwapiti}{294.79\%\xspace}

\newcommand{\improvedCoverageTwoONkanboardVSevocrawl}{138.22\%\xspace}

\newcommand{\improvedCoverageTwoONkanboardVSzap}{83.45\%\xspace}

\newcommand{\improvedCoverageTwoONkanboardVSyura}{164.91\%\xspace}
\newcommand{\improvedCoverageVsAllOnkanboard}{57.32\%\xspace}

\newcommand{\improvedCoverageTwoONwordpressVSblackostrich}{13.56\%\xspace}

\newcommand{\improvedCoverageTwoONwordpressVSarachni}{33.95\%\xspace}

\newcommand{\improvedCoverageTwoONwordpressVSwapiti}{205.28\%\xspace}

\newcommand{\improvedCoverageTwoONwordpressVSevocrawl}{22.93\%\xspace}

\newcommand{\improvedCoverageTwoONwordpressVSzap}{47.78\%\xspace}

\newcommand{\improvedCoverageTwoONwordpressVSyura}{36.83\%\xspace}
\newcommand{\improvedCoverageVsAllOnwordpress}{3.33\%\xspace}

\newcommand{\improvedCoverageTwoONoscommerceVSblackostrich}{48.16\%\xspace}

\newcommand{\improvedCoverageTwoONoscommerceVSarachni}{322.65\%\xspace}

\newcommand{\improvedCoverageTwoONoscommerceVSwapiti}{1442.41\%\xspace}

\newcommand{\improvedCoverageTwoONoscommerceVSevocrawl}{39.26\%\xspace}

\newcommand{\improvedCoverageTwoONoscommerceVSzap}{1103.26\%\xspace}

\newcommand{\improvedCoverageTwoONoscommerceVSyura}{43.65\%\xspace}
\newcommand{\improvedCoverageVsAllOnoscommerce}{9.67\%\xspace}

\newcommand{\improvedCoverageTwoONleantimeVSblackostrich}{9.80\%\xspace}

\newcommand{\improvedCoverageTwoONleantimeVSarachni}{37.15\%\xspace}

\newcommand{\improvedCoverageTwoONleantimeVSwapiti}{28.17\%\xspace}

\newcommand{\improvedCoverageTwoONleantimeVSevocrawl}{9.62\%\xspace}

\newcommand{\improvedCoverageTwoONleantimeVSzap}{8.62\%\xspace}

\newcommand{\improvedCoverageTwoONleantimeVSyura}{51.97\%\xspace}
\newcommand{\improvedCoverageVsAllOnleantime}{-7.84\%\xspace}

\newcommand{\improvedCoverageTwoONpiwigoVSblackostrich}{117.24\%\xspace}

\newcommand{\improvedCoverageTwoONpiwigoVSarachni}{107.32\%\xspace}

\newcommand{\improvedCoverageTwoONpiwigoVSwapiti}{962.42\%\xspace}

\newcommand{\improvedCoverageTwoONpiwigoVSevocrawl}{236.10\%\xspace}

\newcommand{\improvedCoverageTwoONpiwigoVSzap}{308.16\%\xspace}

\newcommand{\improvedCoverageTwoONpiwigoVSyura}{198.66\%\xspace}
\newcommand{\improvedCoverageVsAllOnpiwigo}{74.89\%\xspace}

\newcommand{\improvedCoverageTwoONnextcloudVSblackostrich}{11.79\%\xspace}

\newcommand{\improvedCoverageTwoONnextcloudVSarachni}{372.82\%\xspace}

\newcommand{\improvedCoverageTwoONnextcloudVSwapiti}{421.68\%\xspace}

\newcommand{\improvedCoverageTwoONnextcloudVSevocrawl}{10.63\%\xspace}

\newcommand{\improvedCoverageTwoONnextcloudVSzap}{41.99\%\xspace}

\newcommand{\improvedCoverageTwoONnextcloudVSyura}{17.12\%\xspace}
\newcommand{\improvedCoverageVsAllOnnextcloud}{-7.16\%\xspace}

\newcommand{\improvedCoverageVSblackostrich}{47.74\%\xspace}

\newcommand{\improvedCoverageVSarachni}{219.34\%\xspace}
\newcommand{\improvedCoverageVSwapiti}{462.67\%\xspace}
\newcommand{\improvedCoverageVSevocrawl}{61.18\%\xspace}
\newcommand{\improvedCoverageVSzap}{222.53\%\xspace}
\newcommand{\improvedCoverageVSyura}{84.20\%\xspace}
\newcommand{\improvedCoverageVSAll}{17.44\%\xspace}

\newcommand{\improvedCoverageVSSota}{44.24\%\xspace}

\newcommand{\improvedCoverageVSbwelements}{32\%\xspace}
\newcommand{\improvedCoverageVSbwinputs}{42\%\xspace}

\newcommand{\improvedCoverageApplicationsScannerFROZEN}{46\%\xspace}
\newcommand{\improvedCoverageVSAllNoDecFROZEN}{16\%\xspace}

\newcommand{\para}[1]{\indent {\bf #1.}}

\begin{document}

\title{\shortname: Client-Centric Web Crawler and Security Scanner}

\makeatletter %
\newcommand{\linebreakand}{%
  \end{@IEEEauthorhalign}
  \hfill\mbox{}\par
  \mbox{}\hfill\begin{@IEEEauthorhalign}
}
\makeatother %

\author{
\IEEEauthorblockN{Eric Olsson}
\IEEEauthorblockA{Chalmers University of Technology \\
and University of Gothenburg}
\and
\IEEEauthorblockN{Benjamin Eriksson}
\IEEEauthorblockA{Chalmers University of Technology \\
and University of Gothenburg}
\and
\IEEEauthorblockN{Adam Doup\'e}
\IEEEauthorblockA{Arizona State University}
\and

\linebreakand

\IEEEauthorblockN{Andrei Sabelfeld}
\IEEEauthorblockA{Chalmers University of Technology \\
and University of Gothenburg}
}

\maketitle

\begin{abstract} %

Black-box web application crawling and scanning play an important role for security testing of web applications.
Yet state-of-the-art scanners fall short of addressing key characteristics of a modern web application: its extreme dynamism and interactivity on the client side.
This paper identifies \emph{\technique} as a key ingredient for scanners to deeply explore modern web applications.
We propose \emph{\shortname}, a client-centric crawler and security scanner.
\shortname incorporates a unique combination of high-level, user-facing feedback channels from the web application to achieve \technique in a black-box crawling loop.
These feedback channels include both novel methods to detect interactable elements and sensibly order UI interactions, and orthogonally using an LLM to solve forms.
In doing so, we demonstrate how to reliably discover and test deep states of modern web applications.
Furthermore, our modular approach and useful abstraction layer can serve as a building block for future scanners.
The evaluation of our approach shows substantial improvements in both code coverage and vulnerability detection over previous work.
Our approach increased average code coverage across applications by at least \improvedCoverageApplicationsScannerFROZEN over any other scanner,
or \improvedCoverageVSAllNoDecFROZEN when compared to the union of all other scanners.
We find XSS vulnerabilities in \xssApplications web applications, while any other scanner finds XSS in up to 2 applications.

\end{abstract}

\section{Introduction}

Black-box security scanning is a natural fit for detecting vulnerabilities in web applications.
These scanners use a black-box web crawler to interact with a web application, iteratively discovering functionality and testing it for vulnerabilities.
Because they require only a running web application and a scanner, black-box scanners are largely independent of the frameworks and languages underlying the web application.

\para{Black-box Scanners and the Web}
Historically, as web applications have incorporated more functionality along with their associated complexity, security scanning techniques were developed in short order to handle these new features.
In the 2000s, ``sitemap'' style black-box crawlers such as Skipfish~\cite{skipfish} and w3af~\cite{w3af} could reliably scan the \emph{Web 1.0}---websites composed of primarily statically linked content.
While supporting basic web application interactions via inputs in request parameters and forms, these scanners mainly focused on following static links by parsing HTML.

However, by the early 2010s \emph{Web 2.0} had evolved to include dynamic \emph{stateful web applications} with multi-step flows that could not be modeled well by the first generation of scanners.
New black-box scanning methods such as the Enemy of the State~\cite{DBLP:conf/uss/DoupeCKV12} responded by incorporating a more complex internal state model.
The additional \emph{client-side functionality} of Web 2.0-style applications also motivated the concurrent development of yet more black-box crawlers such as CrawlJax~\cite{DBLP:conf/icwe/MesbahBD08} and \jak~\cite{DBLP:conf/raid/PellegrinoTBR15}, which incorporated support for either AJAX or broader JavaScript applications.

\para{New Scale of Complexity}
In the decade since this second generation of black-box crawlers, the characterization of Web 2.0 as consisting of stateful web applications with rich client-side functionality has remained the same.
However, the \textit{scale of this complexity}, in terms of the dynamic statefulness and interactive client-side functionality of these applications, has exceeded the ability of these black-box scanners to handle.
While
``software is eating the world''~\cite{softwareeatingworld},
we are now in an era where
\emph{web applications eat software}.
Software services that were once offered as desktop applications can now be accessed through web interfaces.
For example, Nextcloud delivers full office productivity suites online through its open-source collaboration platform.

These web interfaces have the same core challenges for crawling, statefulness, and client-side interactivity, but at a new scale due to the inherent complexity of the functionality they now include.
We refer to
web applications with this scale of dynamism and client-side interactivity as \textit{modern web applications}.
Scanning these modern web applications requires that the scanner either be able to navigate multi-step stateful flows, or compose complex structured inputs to backend services that might result from the composition of multiple user steps in the actual application.
Without this, the \textit{deep} behavior of these web applications stays hidden and untested for vulnerabilities.
The key limitation is not merely that current scanners miss individual buttons or fail individual forms.
Rather, they often observe and manipulate the application at abstractions different from those exposed to users: URLs, request parameters,
HTML tag names, or event listeners instead of the actual rendered interface and the valid actions that it enables.

\para{Insufficient Current Approaches}
In practice, black-box scanners only discover a fraction of the functionality and endpoints of these web applications.
Some portion of the application is never tested because discovering and executing a correct sequence of client-side actions is unlikely.
Scanners that avoid client-side actions and instead work directly with server-side requests are also unlikely to succeed, as the problem then transforms to either handling multiple stateful requests, or generating highly structured network inputs.
As a result, current black-box approaches are unlikely to explore these applications deeply.

\para{Interaction strategy and deep states}
Constructing an \textit{interaction strategy} that can navigate the client-side interface of modern web applications to explore their \textit{deep} states has been the elusive target of prior works~\cite{DBLP:conf/dimva/DoupeCV10, DBLP:conf/sp/ErikssonPS21}.
Recent work on \evocrawl~\cite{DBLP:conf/ndss/GuoKLL25} and
\yurascanner~\cite{DBLP:conf/ndss/StafeevRSKP25} makes progress on this goal,
leveraging evolutionary search and LLM task-driven scanning, respectively.
Yet, fundamental challenges of achieving tight interaction with the web application remain open.

\para{Approach}
We identify \emph{\technique} as a key ingredient for scanners to deeply explore modern web applications.
Our approach has the aim of generating actions in a fashion \textit{similar to a human}.
Therefore, we focus on interacting with the application's client-side interface.
This requires the scanner to be aware of how the web application presents itself in the browser: in other words, the signals from high-level, user-facing feedback channels that normal users consume.
Under \emph{\technique},  high-level feedback should be used to inform which actions the scanner performs.
Furthermore, these actions should
be performed as they would be in a browser.
Rather than mainly interacting
through
low-level URLs, network requests, or event listeners,
we instead develop a scanner focused on valid client-side \textit{actions}.

This scanner, \textit{\shortname}, implements a black-box crawling approach that can interact with complex client-side applications by integrating \textit{three pillars of interactivity}: (1) detecting interactable elements, (2) ordering UI interactions, and (3) complex form solving using LLMs.
We develop novel methods for the first two pillars while adopting \yurascanner's approach to form solving.
All of these improvements to specific components are enabled by adding high-level, user-facing feedback channels to the black-box crawler, as shown in \Cref{fig:architecture_diagram}.

\begin{figure}[tb]
	\centering
	\includegraphics[width=0.4\textwidth]{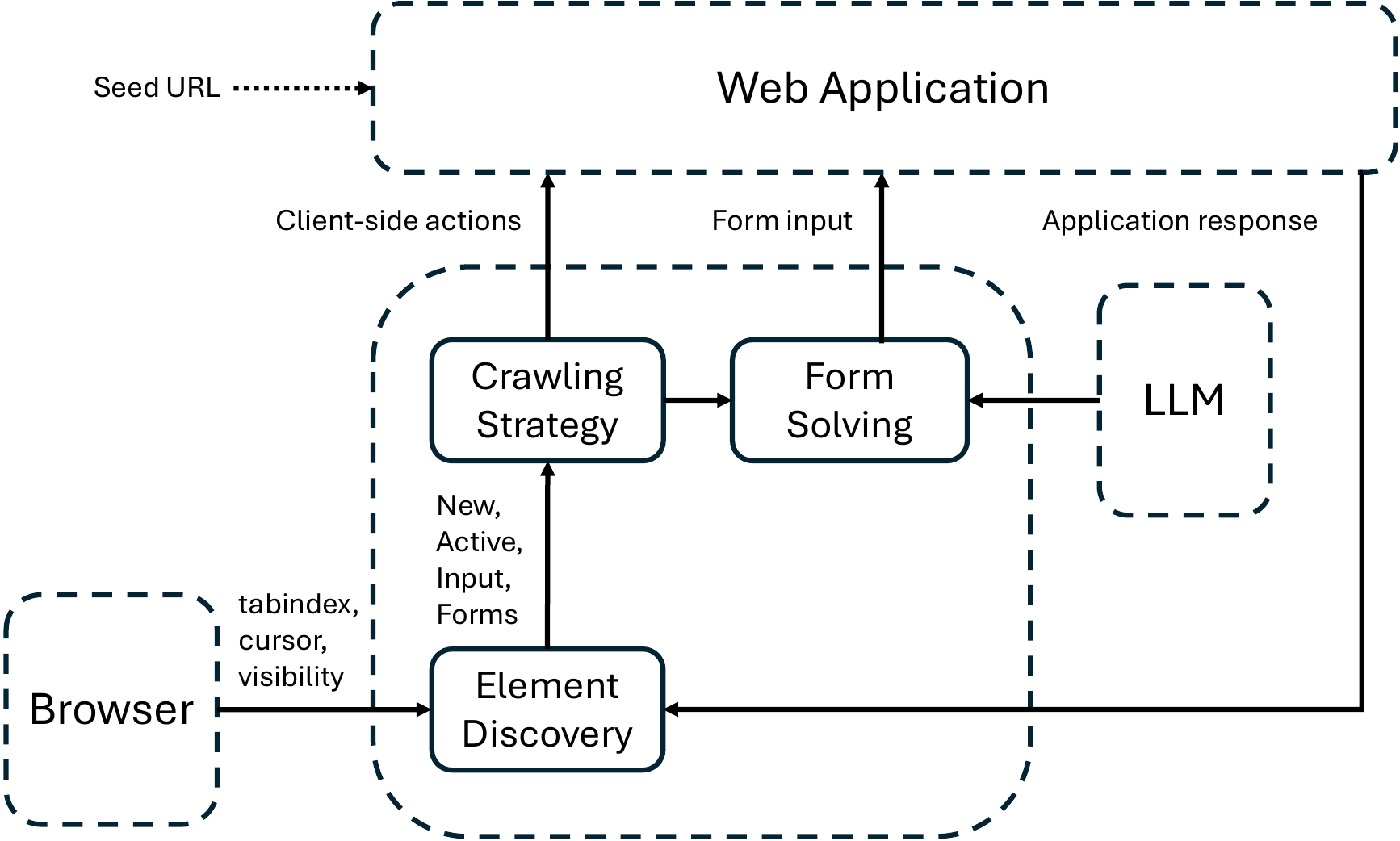}
	\caption{
		The black-box crawling loop of \shortname.
		Starting at a seed URL, we discover interactable elements on the page with user-facing signals from the browser.
		Our crawling strategy selects a valid client-side action to take, using high-level signals about elements.
		If a form is selected, form solving queries an LLM for appropriate input.
		Form inputs and other actions are executed on the web application.
	}
	\label{fig:architecture_diagram}
\end{figure}

Recent black-box crawlers and LLM-assisted scanners address related parts of this problem from different directions.
\evocrawl improves the ordering of UI interactions, and \yurascanner demonstrates how LLMs can support task-oriented scanning and form solving.
\shortname takes a complementary direction: it redesigns the black-box crawling loop around user-facing browser feedback.
The resulting scanner combines framework-agnostic interactable element discovery, interaction ordering, and LLM-based form solving in a recurrent crawler.
This combination allows \shortname to discover new application states and reach in-depth functionality, especially in applications where later actions are only exposed after earlier interactions.
In contrast, single-pass task-solving approaches such as \yurascanner can be limited when effective exploration requires repeatedly discovering, acting on, and revisiting newly exposed states (see \Cref{appendix:task_solvers}).
Compared to both \yura and \evocrawl, our approach alone has general interactable element discovery regardless of client-side framework.
We discuss in detail the differences between these systems throughout \Crefrange{sec:method}{sec:related_work}.
Our method provides a complementary building block to improve the performance of these and future work (see \Cref{sec:future_work}).

\para{Evaluation}
We evaluate crawling performance on \totalApplications open-source web applications compared to \totalScanners black-box scanners.
We select these web applications based on their popularity and inclusion in prior work (see \Cref{tab:web_applications}).
While we characterize these as complex modern applications, arguably ``classic'' applications such as \hotcrp are included.
Yet, \hotcrp is still far from solved (see \Cref{sec:state_depth}).

Our approach increased average code coverage across applications by at least \improvedCoverageApplicationsScannerFROZEN{} over any other scanner,
or \improvedCoverageVSAllNoDecFROZEN{} when compared to the \textit{union} of all other scanners.
Our approach also finds \totalXSS XSS vulnerabilities across \xssApplications of the \totalApplications applications, while \othersXSS are found by the other scanners in \othersXSSApplications applications.
This breadth is important: while other scanners can find many vulnerabilities in one application, any individual scanner finds XSS in at most 2 applications, whereas \shortname finds XSS across \xssApplications applications.

We also perform additional evaluations investigating our scanner and its performance,
by analyzing the application states it reaches,
its client-side coverage,
the non-semantic interactable elements it discovers,
and removing components of it in two ablation studies.

\para{Contributions}
We offer the following contributions:
\setitemize{noitemsep,topsep=0pt,parsep=0pt,partopsep=0pt,leftmargin=*}
\begin{itemize}
\item We develop a novel method that can drive black-box scanners to test deeper program states in modern web applications by using high-level feedback to generate valid client actions and inputs in \Cref{sec:method}.
\item We implement our method into \shortname, a fully black-box crawler and XSS scanner, using an LLM for form solving.
\item We evaluate our black-box scanning method as well as \totalScanners others on \totalApplications web applications in  \Cref{sec:evaluation}.
\shortname finds \totalXSS XSS vulnerabilities in \xssApplications applications.
\item We analyze our findings and share insights for future works in \Cref{sec:analysis}.
\end{itemize}

\section{Challenges}
\label{sec:challenges}

Exploring modern web applications from a black-box perspective remains a difficult task for fully automated systems.
One way to improve scanners is to lift the web application interaction abstractions to be at the same level as actual users.
Taking cues from user-facing signals from the application, and performing valid actions with the client-side interface can allow a scanner to navigate stateful workflows and discover deep states.
Therefore, integrating \technique into a scanner's \emph{interaction strategy} is vital.
We identify interaction strategy as the key overarching challenge and decompose it into three subchallenges.

\para{Interaction Strategy}
\label{sec:black-box}
Black-box crawlers %
must first interact with a web application to discover the endpoints or functionality through an iterative process.
Stafeev and Pellegrino distinguish black-box crawlers by their navigation strategies, used to decide what page to visit and in what order, and page similarity methods,
used to identify duplicate application states \cite{DBLP:conf/uss/StafeevP24}.

Rather than modifying only those components, we change \textit{the whole crawling loop} to function at a more suitable abstraction level for our goal of mimicking normal web application users.
Rather than use low-level URLs, network requests, or JavaScript event listeners, we use high-level valid client-side actions.
Furthermore, user-facing feedback informs users of how to interact with applications.
While an input can be implemented in many different ways, high-level properties of these elements, such as their styling, focus, and other behavior, are the mechanisms web applications use to indicate interactability to their users.
Therefore, our scanner selects client-side actions based on these feedback channels in the browser.
We thereby improve a scanner's ability to interact with and find vulnerabilities deep in modern web applications.

Modern web applications are stateful on both the server- and client-side, and they have complex series of client-side interactions.
Furthermore, these rich client-side interactions are often multi-step:
different options are presented depending on the previous client-side interactions.
While prior black-box scanners can theoretically discover new vulnerable application behavior despite it being locked behind multiple stateful interactions, this is unlikely.

We focus on improving the interaction strategy of black-box crawlers to handle modern web applications.
We realized that prior work, when using aspects of the web application to determine a navigation strategy, i.e., CSS, HTML, and JavaScript, were not using these resources in such a way that the behavior of the rendered DOM was accurately inferred or modeled by the scanner.
We identify three main challenges in adapting a scanner's interaction strategy---(1) \emph{detecting interactable elements}, (2) \emph{ordering interactions}, and (3) \emph{generating inputs for forms}.

\para{Detecting Interactable Elements}
A major source of complexity in modern web applications for scanners is from the client-side JavaScript code.
While the client-side code produces links, buttons, text fields, and other interactable elements that are designed for users to easily identify and interact with, this same task challenges automated scanners. %
Part of the challenge is the lack of HTML semantic meaning applied to interactable elements.
Instead of using the traditional \texttt{a} element or \texttt{button} element, which convey clickable HTML semantics, modern web applications can (and often do) use semantically meaningless (in terms of interactability) \texttt{label} elements or custom tag names,
often due to the frameworks used.
Such elements will not be found by crawlers, including \evocrawl~\cite{DBLP:conf/ndss/GuoKLL25}, that rely on specific HTML tags.
Furthermore, approaches that rely on a fixed set of HTML tags are \emph{fundamentally static} and must be constantly updated to the changing web.

Apart from the semantic meaning of tags, JavaScript event listeners (e.g., \textit{onclick}), can also be used to infer if an element is interactable.
In the simplest case, a static attribute sets the event listener. %
A limitation of state-of-the-art scanners,
including \yura~\cite{DBLP:conf/ndss/StafeevRSKP25} and \bw~\cite{DBLP:conf/sp/ErikssonPS21},
is detecting elements with dynamically assigned event handlers or custom event handling.
This is a common pattern as many JavaScript frameworks (e.g., jQuery) have custom event handling and capture
all events on parent elements (or even the DOM body), which divorces the event listener from the relevant element.
From the scanner's perspective, an element will not have any event listeners, obscuring its interactability.

\para{Order of Interactions}
In addition to correctly modelling which elements can be interacted with, as well as the type of interaction, e.g., click, type text, drag-and-drop, etc., the order of interactions also matters.
For example, a page might have a button that opens a form in a modal overlaid HTML popup.
For the user, it is clear that the form can now be interacted with and that the
covered elements behind the form cannot.
However, a scanner that simply analyzes the HTML and DOM will not recognize the difference between the now interactable form and non-interactable elements behind it and will perform actions in the wrong order.
This will close the form prematurely, losing a chance to interact with it in the correct state.
Alternatively, trying to interact with the elements behind the form will have no effect in their non-interactable state.

Another aspect that complicates this challenge is that a scanner could even click on hidden buttons, as JavaScript allows such behavior.
However, the hidden button receiving the click will likely not have its intended effect in the application.
As such, scanners not only have to find all interactable elements, but also must determine which are currently interactable given the client-side state.
This challenge means that scanners that support client-side actions, such as \jak~\cite{DBLP:conf/raid/PellegrinoTBR15} and \bw~\cite{DBLP:conf/sp/ErikssonPS21}, which might interact with elements in the correct order, still usually fail to access deep program states as they pick elements at random without prioritizing \emph{currently interactable elements}.
While the page abstraction in \yura~\cite{DBLP:conf/ndss/StafeevRSKP25} considers visibility of elements, the scanner does not prioritize
interaction with text elements. Consequently, it might fail to handle UI forms that require text input followed by clicks.

\para{Form Solving}
The final challenge we identified in achieving \technique is that, even after a scanner can detect all interactable elements and understand the proper order of interactions, a scanner will still encounter complex forms in modern web applications.
Thus, a scanner must be able to provide valid inputs to reach deep program states.

Some examples of the inputs for form validation (which can be either performed partially or completely by the browser, client-side JavaScript code, or server-side code) include valid emails, numbers, URLs, zip codes, and addresses.
While HTML can provide semantic attributes for the browser to enforce---that could help the scanner determine input requirements---such annotations are optional.
Examples include the \texttt{type} attribute on the input element and the more general regex-validation attribute \texttt{pattern}.
When validation is performed client-side and these annotations are supplied, black-box crawlers such as \blackostrich~\cite{DBLP:conf/ccs/ErikssonSMRS23}, can often solve their validation constraints and provide a matching input.

However, HTML forms are as diverse as the web itself, and vary in their structure and validation.
Therefore, many forms only explain the expected type of data in descriptive text for the intended user, making it difficult for scanners to automatically infer it.
A more fundamental problem is that server-side code, invisible to the scanners, might perform data validation on the input.
This web application logic can enforce validations that HTML semantics do not support.
For example, required relations between inputs (e.g., in user registration forms the ``password'' and ``confirm password'' combination must be the same), or required relations between inputs and values in a database (e.g., inputs that must be unique across the application, such as ``username'' and ``email'').
The only hint of such validation can be in error messages, which may further lack any explicit relationship to the specific erroneous element/s.

\section{Method}
\label{sec:method}
In each of the prior challenges, the low-level abstractions of previous methods fail to reliably discover interactable elements, choose their ordering, and provide form inputs in modern web applications.
We propose \shortname, a black-box crawling approach driven by \technique, overviewed in \Cref{fig:architecture_diagram}.
\shortname contributes novel methods for detecting interactable elements and ordering UI interactions.
It also uses LLMs to solve forms, following the approach pioneered by \yurascanner.
In doing so, we aim to more reliably discover and test deep states of modern web applications by operating at a higher abstraction level.

\subsection{Motivating Example}
To exemplify the increased complexity of modern web applications we present the following realistic example based on
challenges seen in the core functionality of
\piwigo and \kanboard.
In a project-management web application,
users can create new tasks
by clicking on a button  \textcircled{\scriptsize 1} to generate a modal pop-up including a form.
The user can then interact with the form \textcircled{\scriptsize 2},
allowing them to fill in the task details \textcircled{\scriptsize 3} and submit.
From the perspective of a scanner, there are multiple challenges (as detailed in Section~\ref{sec:challenges}).

\begin{figure}[tb]
	\centering
	\includegraphics[width=0.5\textwidth]{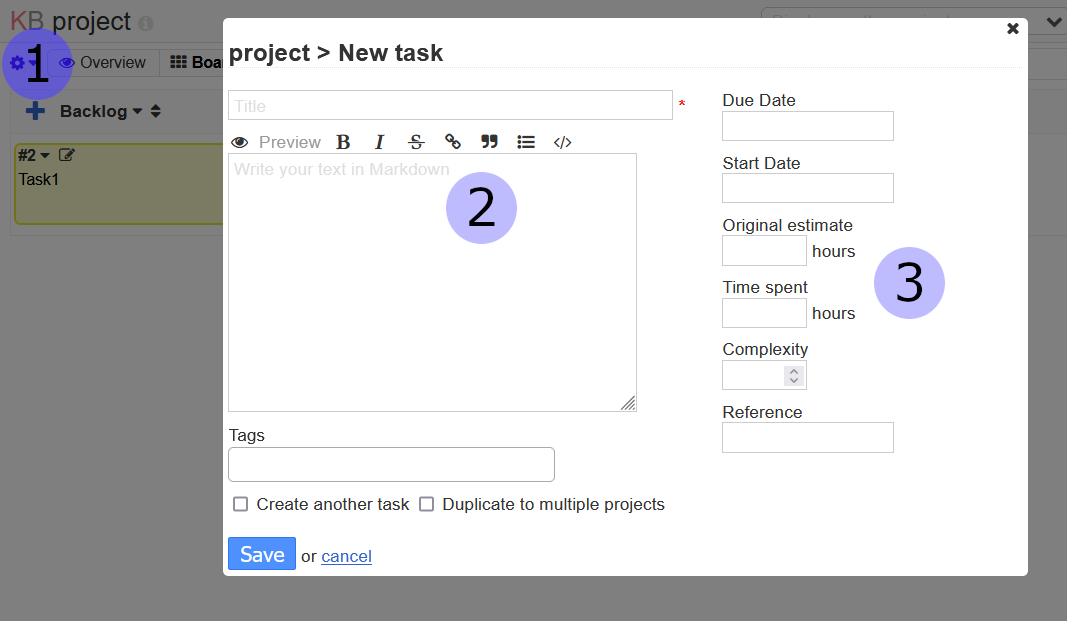}
	\caption{Web application flow with scanning challenges. The scenario is based on two real-world application workflows.}
	\label{fig:motivating_example}
\end{figure}

Clicking on the button (Step \textcircled{\scriptsize 1} in \Cref{fig:motivating_example}) is challenging because it is not actually a \texttt{button} HTML element but instead a \texttt{label} (as is the case in \piwigo), %
and only the styling makes it visually appear to be a button.
Furthermore, this \texttt{label} does not have any event listeners directly on the element, as the application instead relies on a global event handler from the popular jQuery library.

After clicking, the form appears, and the next challenge for the scanner is to prioritize these newly visible and interactive elements in step \textcircled{\scriptsize 2}.
If this is not taken into account, the scanner might try to interact with the blocked elements behind the modal, which will not trigger their event listeners.
Clicking outside the modal causes the form to disappear, preventing future form interactions.

Lastly, the form itself contains further data validation challenges as the user must enter a numeric ``number of hours'' in step  \textcircled{\scriptsize 3}.
This validation requirement is not in the standard HTML semantics, such as \texttt{type="number"} or regex patterns, that would otherwise indicate that the values must be numeric.

\subsection{Client-side Crawling}

We design our tool \shortname to implement \technique, driving the web application into deep program states by reliably discovering and handling modern web application interactions.
\shortname is built on top of the open-source \bw~\cite{DBLP:conf/sp/ErikssonPS21} scanner,
with tailored enhancements to the black-box crawling loop to enable \technique. This is accomplished
by identifying interactable elements, prioritizing element interactions, and using an LLM to fill in forms in the scanner.

\subsubsection{Interactable Element Discovery}

As discussed in Section~\ref{sec:challenges}, a scanner's ability to discover new web application behavior (and test that behavior) depends on it being able to identify the interactable elements on any given page.
Prior scanners generally approach this task by either:
(1) using predefined lists of interactable elements, usually based on the HTML semantics~\cite{arachniUiForm} such as \evocrawl, or
(2) identifying elements with associated event listeners and, possibly, hooking these event listeners such as \yurascanner.
While not commonly used in security scanners, Firefox~\cite{firefoxcode} uses a (3) third option by applying framework-specific static analysis of JavaScript to identify custom event listeners from libraries.

However, these approaches do not generalize, either by:
(1) not including custom element types,
(2) failing to properly associate an event listener to a specific element, or
(3) failing to implement
framework-specific static analysis of other libraries.

In this work, while we use an initial list of semantic HTML input types (\texttt{a}, \texttt{button}, \texttt{form}, and \texttt{input}), our approach extends this continually with elements that indicate their interactability at runtime.
This allows our method to handle step \textcircled{\scriptsize 1} of the motivating example by correctly identifying the \texttt{label} acting as a button.
\evocrawl and \yurascanner do not identify this interactable element.

We rely on the observation that application developers, JavaScript frameworks, and web browsers all aim to guide their human users with various hints of how they can use the web applications. %
Providing these hints is also useful for other
accessibility clients or extensions, such as screen readers.
However, some of these interactability hints cannot be generalized. %
For instance, a user might infer interactability based on the color scheme. %
Therefore, we identify further interactable elements, regardless of HTML type or framework, by inspecting the cursor behavior and tab index of elements.

By dynamically checking the \texttt{cursor} property and \texttt{tabindex} of an element we can draw insight into the developer's intended element semantics.
Specifically, we check for the \texttt{pointer} value or a nonnegative tabindex value to determine if an element is clickable.
According to the CSS specification~\cite{css-pointer}, the \texttt{pointer} cursor should indicate a link.
However, many single-page applications and frameworks, including React, Vue, and Angular~\cite{todomvc}, use the HTML \texttt{a} tag (link) to perform client-side actions
by using event listeners on the tag and using fragment URLs or the \texttt{javascript:} pseudo protocol.
The tabindex~\cite{tabindex} attribute is another method developers can use to indicate that an element is focusable, and often also interactable.
This tabindex attribute plays an important role in helping users of assistive technologies navigate websites.
Similarly to \texttt{pointer}, we use the \texttt{text} value of the cursor to identify an element that supports text input.
This is particularly useful for complex rich-text editors that rely on \texttt{div}s for user inputs. %

We also prune the list of interactable elements to prioritize active elements by inspecting their visibility at runtime.
Elements hidden underneath others cannot be clicked or otherwise interacted with, either from the perspective of a user or Selenium automation driving a browser.
An alternative approach could use JavaScript to force hidden element interactions.
However, we argue that this can lead to unintended client-side states, which we mostly try to avoid
(other than injecting XSS payloads).
Therefore, the interactable element set includes only currently visible and active elements.
To avoid interacting multiple times with the same element, we ignore elements that are already inside a
semantic element. For example, we will not click on a \texttt{span} inside a \texttt{button}.

As an optimization, we also prioritize elements that are in the current DOM (i.e., have a valid Selenium reference).
This allows the scanner to achieve a deeper crawl of the client-side, as opposed to a shallow crawl across the entire application, where elements from different URLs are selected.
When picking an element that is not currently active, the scanner will retrace its previous steps. %
Overall interactable element discovery is described by \Cref{alg:discovery}.

Compared to prior scanners, which are limited to parsing HTML for specific tags, or hooking JavaScript event listeners, our method does not need to model the DOM, and instead uses it directly as rendered by the browser.

\begin{algorithm}[tb]
	\caption{Algorithm to discover interactable elements.
	}

	\begin{algorithmic} %
		\State $semanticElements \gets \texttt{<a>, <button>, <input>}, ...$
		\State $immersiveElements \gets \texttt{pointer, text, tabIndex}$
		\State $elements \gets semanticElements \cup immersiveElements$

		\State $interactables \gets \{\}$
		\For{$element \in elements$}
		  \State $\{x, y\} \gets element.coordinates() $
		  \State $topElement \gets getElementBy(x,y) $

		  \If{$element == topElement$}
		  	\State $interactables.add(element)$
		  \EndIf
		\EndFor

	\end{algorithmic}
	\label{alg:discovery}
\end{algorithm}

\subsubsection{Crawling Strategy}

After discovering the valid interactable elements on a page, our scanner must select an order to interact with these elements.
We attempt to prioritize client-side interactions for exploration, aiming to reach deep modern web application states.
However, we also need to balance client-side exploration interactions with
opportunities to inject exploit payloads, such as form submission and input elements.
Finally, the crawler must handle classic exploration with normal static links.

Randomization is also important for coverage~\cite{DBLP:conf/uss/StafeevP24}.
To accommodate this our method can, with low probability, randomly pick edges.
While this can seem counterintuitive, randomness helps a scanner get unstuck, as there are cases where a scanner can otherwise continuously prioritize a similar element.

Therefore, we design a two-phase approach to crawling.
The crawler tracks actions possible from states in the web application in a directed graph, where edges and nodes represent actions and states.
Actions include following links and interacting with client-side elements and forms.
As we crawl, %
we also prioritize inputs where a payload can be inserted.
This crawling strategy allows our scanner to handle step \textcircled{\scriptsize 2} of the motivating example by prioritizing the correct visible and interactive form elements.

In the first phase, the scanner performs a short shallow scan, prioritizing finding
new linked pages
until either a threshold of discovered links is met, or the scanner cannot find any new links.
After that phase, the scanner enters its main phase where client-side interactions and payload injections are prioritized.
The general prioritization follows three categories (in this order): (1) \emph{new} actions, (2) \emph{active} elements on the current page, and (3) \emph{input} elements.

Because we prioritize active elements, the algorithm will act similarly to a DFS strategy.
For example, a form (``input'') that is on the current page (``active'') and not taken before (``new'') will have a high priority.
To avoid getting stuck in one part of an application (such as the infinite calendar module in WackoPicko~\cite{DBLP:conf/uss/DoupeCKV12}), we have two mechanisms to divert the scanner.
First, if enough iterations have passed since any new edges have appeared, an active link is chosen randomly instead.
If there are also no new forms, we pick a random link with a low probability (5\% in the evaluation).

Unlike scanners that separate exploration and exploitation, such as \zap or \yurascanner, our method aims to always pick the best option when presented, and does not need to terminate.

\subsection{LLM-based Input Generation}

As forms are one of the main ways to provide complex user input to a web application, determining a set of valid inputs that allow a form to be submitted is critical to a web scanner.
The constraint validating the content of the numeric ``number of hours'' in step \textcircled{\scriptsize 3} of the motivating example is in custom JavaScript code rather than a regex pattern.
Prior to \yurascanner's application of LLMs to form solving,
previous methods could not expect to reliably detect a pattern while only supporting a small set of possible implementations.

\para{Large Language Models}
We leverage the single- or few-shot ability of LLMs to provide acceptable form inputs, as pioneered by \yurascanner,
and adopt their method using few-shot examples~\cite{DBLP:conf/nips/BrownMRSKDNSSAA20} and summarizing some important elements of the form.
An LLM can reasonably infer that an input described by ``number of hours'', from step \textcircled{\scriptsize 3} of the motivating example, should be numeric and in an appropriate range.

\para{Implementation}
\label{sec:llm-implementation}
When the scanner finds a form it provides the form's HTML to the LLM form solving module.
The LLM is prompted to solve this form, producing a list of (selector, value) pairs to indicate what values should be input to which elements.
The scanner then submits the form with these values.
Besides the form HTML in the prompt, we also specifically include all input elements, those marked required, and those with regex patterns.
Additional details about the implementation and prompt are included in \Cref{sec:llm_form_details}.

\colorlet{cl1}{red!20!white}
\colorlet{cl2}{brown!20!white}
\colorlet{cl3}{blue!20!white}

\para{Other Form Adaptations}
\label{llm-crawl-subprompts}
Finally, before submitting a form we prompt an LLM to check if the form is ``dangerous'', and could change the current username or password, to avoid breaking the application.
We also use an LLM to determine the correct submit button in a form, as buttons can
have wildly different semantics including canceling the submission.
We also implement handling for \emph{drag-and-drop} functionality.
If a form supports file upload based on its \texttt{enctype}, we simulate a drag-and-drop to all form elements from  a standard list of prepared mock files.

\subsection{Implementation}

We implement \shortname in Python extending the \bw scanner~\cite{DBLP:conf/sp/ErikssonPS21}.
We implement LLM-facing code in LangChain to be able to easily swap the underlying model. %
While we have tested this LLM code with versions of Google's Gemini, OpenAI's GPT, and local Ollama models, we conduct our evaluation with Gemini, using the 2.5 Flash model, and a temperature of 1.7.
Details about the model, temperature, tokens, and cost can be found in \Cref{appendix:llm_details}.

\section{Evaluation}
\label{sec:evaluation}
We evaluate the entire scanner's performance, including both improved
client-side crawling and LLM form solving, in the task of exploring a web
application and discovering XSS vulnerabilities.
This scanner evaluation is performed on a set of open-source web applications (\Cref{sec:webapp_crawl_eval}) and presented in \Cref{sec:eval_results}.
Based on this we perform a more detailed comparison with recent state-of-the-art scanners in \Cref{sec:recent_sota}.
We perform a case study with \yurascanner in \Cref{sec:repeated_evaluation} where we repeat the evaluations
and investigate how results differ across two scanning sessions. %
In \Cref{sec:state_depth} we include a novel scanner-agnostic data-driven state-depth analysis.
\Cref{sec:clientside_coverage} presents a new method for comparing client-side coverage in scanners
and compares our method to \blackostrich.
In \Cref{sec:nonsemantic_interactable} we evaluate the amount of non-semantic interactable elements our scanner finds.
Finally, in \Cref{sec:ablation_study_summary} we summarize the results of two ablation studies that remove either the LLM form solving or
the element detection components, to evaluate the individual contribution from each of the
components of our method.
These ablation studies are fully described in \Cref{sec:ablation_study}.

\subsection{Web Application Crawl Evaluation}
\label{sec:webapp_crawl_eval}
The overall goal of this work is to better explore web applications from a black-box perspective, by successfully navigating client-side interactions %
to reach new states.
By exploring more states in these web applications, we also hope to find new vulnerabilities.
To gather relevant performance metrics and analyze vulnerabilities,
we compare our crawler's performance to other relevant state-of-the-art black-box techniques on a set of local open-source web applications.

Similar to other
scanner evaluations~\cite{DBLP:conf/sp/ErikssonPS21, DBLP:conf/ndss/StafeevRSKP25, DBLP:conf/ndss/GuoKLL25, DBLP:conf/uss/0001EDS24, DBLP:conf/raid/PellegrinoTBR15}, we primarily use single runs of (scanner, web application) pairs to compare to prior work.
We note that while our goal of improving code coverage to find new vulnerabilities is shared with almost all prior scanning work, not all code is actually security-relevant and discuss this in \Cref{sec:analysis_intended}.

\subsubsection{Open-source Web Applications}
\label{sec:web_apps}

We choose \totalApplications open-source web applications for this evaluation:
\dokuwiki,
\hotcrp,
\kanboard,
\leantime,
\nextcloud,
\oscommerce,
\piwigo,
\tinyfilemanager,
and
\wordpress.
We describe these applications and their versions in \Cref{appendix:web_apps}.
These applications are chosen based on their popularity and inclusion in related work. %
Our selection criteria were collecting PHP web applications from prior work, removing older applications without client-side functionality (e.g., MyBB), then adding additional real and available applications with client-side functionality.

We limit the evaluation to PHP applications to
allow for uniform collection of server-side coverage.
In \Cref{tab:frontend-frameworks}, we describe which front-end JS frameworks are used in these applications.
While jQuery is used across this set, they also rely on Bootstrap, React, and Vue.js.
We strive to include the latest version of modern applications. %
With the exception of \tinyfilemanager, a popular single-page application, all other applications are also included in other evaluations.
We make slight modifications to these applications to level the playing field for all scanners (standardized by Doup\'e et al.~\cite{DBLP:conf/dimva/DoupeCV10}),
mainly with respect to authentication; details are provided in \Cref{appendix:web_apps}.

\subsubsection{Compared Black-box Scanners}

We evaluate our approach against both state-of-the-art black-box academic and open-source scanners.
This set of \totalScanners scanners consists of:
\arachni~\cite{arachni}, \blackostrich~\cite{DBLP:conf/ccs/ErikssonSMRS23}, \wapiti~\cite{wapiti}, \zap~\cite{owaspzap}, and the latest \evocrawl~\cite{DBLP:conf/ndss/GuoKLL25} and \yurascanner~\cite{DBLP:conf/ndss/StafeevRSKP25}.
When possible, we configure the scanners to only detect XSS vulnerabilities.
We exclude \bw from this comparison, as \blackostrich directly extends that scanner with regex-based input validation support.
In the case of \yurascanner, we configure it with the same Gemini LLM as \shortname to allow for a fair comparison.
All scanners are run for 8 hours for each application.

\subsubsection{Metrics}
The main metrics we focus on are code coverage and the number of XSS vulnerabilities found.
For vulnerabilities, we present the number of reports from each scanner after we manually verify reports to filter out false positives.
Some scanner's verifications of injections, including statically searching for \texttt{script} tags or not using unique IDs for payloads, can result in false positives.
All vulnerabilities are manually deduplicated based on shared server-side code.

In this study, we define coverage in terms of unique lines of code (LoC) executed on the server-side
collected using xDebug in PHP, similar to prior work~\cite{DBLP:conf/uss/StafeevP24, DBLP:conf/ndss/GuoKLL25, DBLP:conf/sp/ErikssonPS21}.
We note that coverage could be measured differently~\cite{DBLP:conf/raid/PellegrinoTBR15, DBLP:conf/uss/StafeevP24, DBLP:conf/ndss/Kang0C22} and discuss this
in \Cref{sec:discussion_metric}.

\subsection{Web Application Crawl Evaluation Results}
\label{sec:eval_results}
While \shortname is able to reach new coverage and program states in the evaluated applications, other scanners can still find things we miss.
For example, numerous error states are missed in  \tinyfilemanager.
However, \shortname is the only scanner in our evaluation able to create a task in \kanboard in our evaluation,
or review a paper in \hotcrp.
\shortname achieves better server-side coverage in the vast majority of cases.
Furthermore, \shortname finds more XSS vulnerabilities, and importantly, across more applications, than all other scanners.
In the following sections, we analyze the results of the local web application crawling evaluation in terms of coverage and vulnerabilities.

\subsubsection{Coverage}
\label{sec:eval_coverage}
In this section, we present the server-side code coverage from running all the scanners on
our set of open-source web applications.
In \Cref{fig:code_coverage,fig:code_coverage_recent_sota}
we plot the comparisons between our scanner and the others, with
the exact numbers presented in \Cref{appendix:coverage_results}, and improvement in \Cref{table:coverage_improvement}.
Our method outperforms the other scanners in 7/\totalApplications of the evaluated applications, indicating that focusing on client-side crawling does achieve deeper
scanning of these applications, even for server-side code.

A notable case where another scanner outperforms us is
\zap on \tinyfilemanager{}.
In this case, \zap{} triggers more errors, such as invalid CSRF tokens,
which we avoid by correctly retracing the form submissions.
We discuss the differences between coverage of \textit{intended code} versus
\textit{error code} in \Cref{sec:analysis_intended}.
Furthermore, in \Cref{sec:discussion_time} we discuss using scanners for deeper, longer scans than commonly used in either this or prior evaluations.

\newcommand{\blackostrichshort}{Black Ost.\xspace}
\newcommand{\tinyfilemanagershort}{TinyFM\xspace}
  \begin{figure*}
  \begin{tikzpicture}
  \begin{axis}[
      reverse legend,
      ybar stacked,
      bar width=6pt,
      enlargelimits=0.05,
          legend style={font=\small, at={(0.155,0.9)}, cells={align=left}},
          legend cell align={left},
          legend entries={\tiny Other scanner, \tiny Common, \tiny Our Scanner},
      ymajorgrids,
      major grid style={gray},
      enlarge x limits={abs=6pt},
      width=1\textwidth,
      height=7cm,
      yticklabel={\tiny\pgfmathprintnumber{\tick}\%},
      cycle list name=mycolorlist,
      symbolic x coords={\oscommerce \texttt{\space\space\space\space\space \wapiti},\oscommerce \texttt{\space\space\space\space\space\space\space\space \zap},\piwigo \texttt{\space\space\space\space\space \wapiti},\hotcrp \texttt{\space\space\space\space\space \wapiti},\hotcrp \texttt{\space\space\space\space \arachni},\nextcloud \texttt{\space\space\space\space\space \wapiti},\hotcrp \texttt{\space\space\space\space\space\space\space\space \zap},\nextcloud \texttt{\space\space\space\space \arachni},\oscommerce \texttt{\space\space\space\space \arachni},\kanboard \texttt{\space\space\space\space\space \wapiti},\piwigo \texttt{\space\space\space\space\space\space\space\space \zap},\wordpress \texttt{\space\space\space\space\space \wapiti},\dokuwiki \texttt{\space\space\space\space \arachni},\kanboard \texttt{\space\space\space\space \arachni},\piwigo \texttt{\space \blackostrichshort},\piwigo \texttt{\space\space\space\space \arachni},\kanboard \texttt{\space \blackostrichshort},\hotcrp \texttt{\space \blackostrichshort},\kanboard \texttt{\space\space\space\space\space\space\space\space \zap},\tinyfilemanagershort \texttt{\space\space\space\space\space \wapiti},\wordpress \texttt{\space\space\space\space\space\space\space\space \zap},\oscommerce \texttt{\space \blackostrichshort},\tinyfilemanagershort \texttt{\space\space\space\space \arachni},\nextcloud \texttt{\space\space\space\space\space\space\space\space \zap},\tinyfilemanagershort \texttt{\space \blackostrichshort},\leantime \texttt{\space\space\space\space \arachni},\wordpress \texttt{\space\space\space\space \arachni},\leantime \texttt{\space\space\space\space\space \wapiti},\dokuwiki \texttt{\space \blackostrichshort},\wordpress \texttt{\space \blackostrichshort},\dokuwiki \texttt{\space\space\space\space\space \wapiti},\nextcloud \texttt{\space \blackostrichshort},\leantime \texttt{\space \blackostrichshort},\leantime \texttt{\space\space\space\space\space\space\space\space \zap},\dokuwiki \texttt{\space\space\space\space\space\space\space\space \zap},\tinyfilemanagershort \texttt{\space\space\space\space\space\space\space\space \zap}},
      xtick=data,
      x tick label style={rotate=90,anchor=east,font=\tiny},
      ]

\node[text=red] at (0, 580 ) {\scriptsize 19};
\draw[fill=white, opacity=1.0,  xscale=0.1] (2, 580) circle[radius=30];
\node at (1, 580 ) {\includegraphics[width=0.75cm]{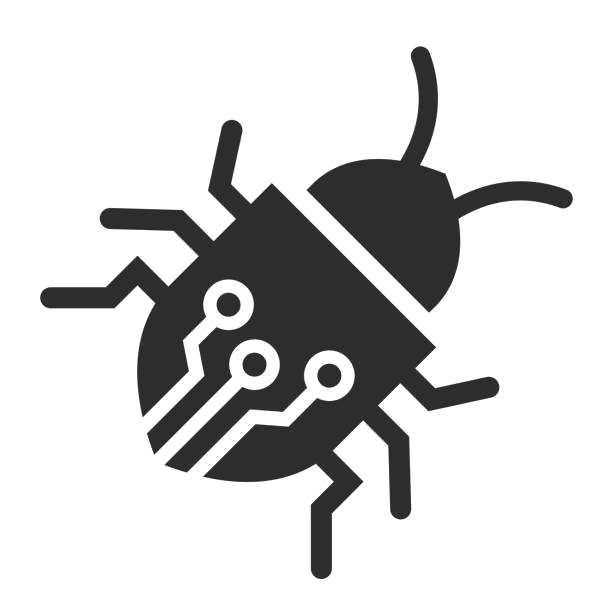}};

\node[text=red] at (10, 580 ) {\scriptsize 19};
\draw[fill=white, opacity=1.0,  xscale=0.1] (102, 580) circle[radius=30];
\node at (11, 580 ) {\includegraphics[width=0.75cm]{bug.png}};

\node[text=red] at (20, 580 ) {\scriptsize 5};
\draw[fill=white, opacity=1.0,  xscale=0.1] (202, 580) circle[radius=30];
\node at (21, 580 ) {\includegraphics[width=0.75cm]{bug.png}};

\node[text=red] at (30, 580 ) {\scriptsize 1};
\draw[fill=white, opacity=1.0,  xscale=0.1] (302, 580) circle[radius=30];
\node at (31, 580 ) {\includegraphics[width=0.75cm]{bug.png}};

\node[text=red] at (40, 580 ) {\scriptsize 1};
\draw[fill=white, opacity=1.0,  xscale=0.1] (402, 580) circle[radius=30];
\node at (41, 580 ) {\includegraphics[width=0.75cm]{bug.png}};

\node[text=red] at (60, 580 ) {\scriptsize 1};
\draw[fill=white, opacity=1.0,  xscale=0.1] (602, 580) circle[radius=30];
\node at (61, 580 ) {\includegraphics[width=0.75cm]{bug.png}};

\node[text=red] at (80, 950 ) {\scriptsize 19};
\draw[fill=white, opacity=1.0,  xscale=0.1] (802, 950) circle[radius=30];
\node at (81, 950 ) {\includegraphics[width=0.75cm]{bug.png}};

\node[text=red] at (90, 950 ) {\scriptsize 1};
\draw[fill=white, opacity=1.0,  xscale=0.1] (902, 950) circle[radius=30];
\node at (91, 950 ) {\includegraphics[width=0.75cm]{bug.png}};

\node[text=red] at (100, 950 ) {\scriptsize 5};
\draw[fill=white, opacity=1.0,  xscale=0.1] (1002, 950) circle[radius=30];
\node at (101, 950 ) {\includegraphics[width=0.75cm]{bug.png}};

\node[text=red] at (110, 950 ) {\scriptsize 3};
\draw[fill=white, opacity=1.0,  xscale=0.1] (1102, 950) circle[radius=30];
\node at (111, 950 ) {\includegraphics[width=0.75cm]{bug.png}};

\node[text=red] at (130, 950 ) {\scriptsize 1};
\draw[fill=white, opacity=1.0,  xscale=0.1] (1302, 950) circle[radius=30];
\node at (131, 950 ) {\includegraphics[width=0.75cm]{bug.png}};

\node[text=red] at (140, 950 ) {\scriptsize 5};
\draw[fill=white, opacity=1.0,  xscale=0.1] (1402, 950) circle[radius=30];
\node at (141, 950 ) {\includegraphics[width=0.75cm]{bug.png}};

\node[text=red] at (150, 950 ) {\scriptsize 5};
\draw[fill=white, opacity=1.0,  xscale=0.1] (1502, 950) circle[radius=30];
\node at (151, 950 ) {\includegraphics[width=0.75cm]{bug.png}};

\node[text=red] at (160, 950 ) {\scriptsize 1};
\draw[fill=white, opacity=1.0,  xscale=0.1] (1602, 950) circle[radius=30];
\node at (161, 950 ) {\includegraphics[width=0.75cm]{bug.png}};

\node[text=red] at (170, 950 ) {\scriptsize 1};
\draw[fill=white, opacity=1.0,  xscale=0.1] (1702, 950) circle[radius=30];
\node at (171, 950 ) {\includegraphics[width=0.75cm]{bug.png}};

\node[text=red] at (180, 950 ) {\scriptsize 1};
\draw[fill=white, opacity=1.0,  xscale=0.1] (1802, 950) circle[radius=30];
\node at (181, 950 ) {\includegraphics[width=0.75cm]{bug.png}};

\node[text=red] at (190, 950 ) {\scriptsize 1};
\draw[fill=white, opacity=1.0,  xscale=0.1] (1902, 950) circle[radius=30];
\node at (191, 950 ) {\includegraphics[width=0.75cm]{bug.png}};

\node[text=red] at (200, 950 ) {\scriptsize 3};
\draw[fill=white, opacity=1.0,  xscale=0.1] (2002, 950) circle[radius=30];
\node at (201, 950 ) {\includegraphics[width=0.75cm]{bug.png}};

\node[text=red] at (210, 950 ) {\scriptsize 19};
\draw[fill=white, opacity=1.0,  xscale=0.1] (2102, 950) circle[radius=30];
\node at (211, 950 ) {\includegraphics[width=0.75cm]{bug.png}};

\node[text=red] at (220, 450 ) {\scriptsize 1};
\draw[fill=white, opacity=1.0,  xscale=0.1] (2202, 450) circle[radius=30];
\node at (221, 450 ) {\includegraphics[width=0.75cm]{bug.png}};

\node[text=red] at (220, 10 ) {\scriptsize 2};
\draw[fill=white, opacity=1.0,  xscale=0.1] (2202, 10) circle[radius=30];
\node at (221, 10 ) {\includegraphics[width=0.75cm]{bug.png}};

\node[text=red] at (240, 950 ) {\scriptsize 1};
\draw[fill=white, opacity=1.0,  xscale=0.1] (2402, 950) circle[radius=30];
\node at (241, 950 ) {\includegraphics[width=0.75cm]{bug.png}};

\node[text=red] at (250, 950 ) {\scriptsize 2};
\draw[fill=white, opacity=1.0,  xscale=0.1] (2502, 950) circle[radius=30];
\node at (251, 950 ) {\includegraphics[width=0.75cm]{bug.png}};

\node[text=red] at (260, 950 ) {\scriptsize 3};
\draw[fill=white, opacity=1.0,  xscale=0.1] (2602, 950) circle[radius=30];
\node at (261, 950 ) {\includegraphics[width=0.75cm]{bug.png}};

\node[text=red] at (270, 950 ) {\scriptsize 2};
\draw[fill=white, opacity=1.0,  xscale=0.1] (2702, 950) circle[radius=30];
\node at (271, 950 ) {\includegraphics[width=0.75cm]{bug.png}};

\node[text=red] at (290, 950 ) {\scriptsize 2};
\draw[fill=white, opacity=1.0,  xscale=0.1] (2902, 950) circle[radius=30];
\node at (291, 950 ) {\includegraphics[width=0.75cm]{bug.png}};

\node[text=red] at (290, 450 ) {\scriptsize 1};
\draw[fill=white, opacity=1.0,  xscale=0.1] (2902, 450) circle[radius=30];
\node at (291, 450 ) {\includegraphics[width=0.75cm]{bug.png}};

\node[text=red] at (320, 950 ) {\scriptsize 2};
\draw[fill=white, opacity=1.0,  xscale=0.1] (3202, 950) circle[radius=30];
\node at (321, 950 ) {\includegraphics[width=0.75cm]{bug.png}};

\node[text=red] at (330, 950 ) {\scriptsize 2};
\draw[fill=white, opacity=1.0,  xscale=0.1] (3302, 950) circle[radius=30];
\node at (331, 950 ) {\includegraphics[width=0.75cm]{bug.png}};

\node[text=red] at (350, 950 ) {\scriptsize 1};
\draw[fill=white, opacity=1.0,  xscale=0.1] (3502, 950) circle[radius=30];
\node at (351, 950 ) {\includegraphics[width=0.75cm]{bug.png}};

\addplot+[ybar] plot coordinates { (\oscommerce \texttt{\space\space\space\space\space \wapiti}, 0.002478294272661936) (\oscommerce \texttt{\space\space\space\space\space\space\space\space \zap}, 0.002478294272661936) (\piwigo \texttt{\space\space\space\space\space \wapiti}, 0.00784313725490196) (\hotcrp \texttt{\space\space\space\space\space \wapiti}, 0.043901769790094664) (\hotcrp \texttt{\space\space\space\space \arachni}, 0.12885184779032788) (\nextcloud \texttt{\space\space\space\space\space \wapiti}, 0.02366338830141241) (\hotcrp \texttt{\space\space\space\space\space\space\space\space \zap}, 0.757348734573787) (\nextcloud \texttt{\space\space\space\space \arachni}, 0.16245993885598664) (\oscommerce \texttt{\space\space\space\space \arachni}, 1.3857546700991454) (\kanboard \texttt{\space\space\space\space\space \wapiti}, 0.08392279103225032) (\piwigo \texttt{\space\space\space\space\space\space\space\space \zap}, 1.2681264641536465) (\wordpress \texttt{\space\space\space\space\space \wapiti}, 0.10368933337793089) (\dokuwiki \texttt{\space\space\space\space \arachni}, 0.30411353572000216) (\kanboard \texttt{\space\space\space\space \arachni}, 0.9566819181175352) (\piwigo \texttt{\space \blackostrichshort}, 1.9628198473575946) (\piwigo \texttt{\space\space\space\space \arachni}, 2.755477584332869) (\kanboard \texttt{\space \blackostrichshort}, 1.6056670602125147) (\hotcrp \texttt{\space \blackostrichshort}, 1.12368699617295) (\kanboard \texttt{\space\space\space\space\space\space\space\space \zap}, 1.4485898421332704) (\tinyfilemanagershort \texttt{\space\space\space\space\space \wapiti}, 2.800546448087432) (\wordpress \texttt{\space\space\space\space\space\space\space\space \zap}, 1.6676269651823195) (\oscommerce \texttt{\space \blackostrichshort}, 2.8382456816284596) (\tinyfilemanagershort \texttt{\space\space\space\space \arachni}, 4.23956931359354) (\nextcloud \texttt{\space\space\space\space\space\space\space\space \zap}, 3.6968971706983504) (\tinyfilemanagershort \texttt{\space \blackostrichshort}, 2.3335621139327385) (\leantime \texttt{\space\space\space\space \arachni}, 3.641991570073762) (\wordpress \texttt{\space\space\space\space \arachni}, 3.8472682785486625) (\leantime \texttt{\space\space\space\space\space \wapiti}, 3.5530652603823336) (\dokuwiki \texttt{\space \blackostrichshort}, 0.26154256738724313) (\wordpress \texttt{\space \blackostrichshort}, 2.142857142857143) (\dokuwiki \texttt{\space\space\space\space\space \wapiti}, 7.837237977805179) (\nextcloud \texttt{\space \blackostrichshort}, 1.9721863716121173) (\leantime \texttt{\space \blackostrichshort}, 4.272441769170374) (\leantime \texttt{\space\space\space\space\space\space\space\space \zap}, 4.860682121143155) (\dokuwiki \texttt{\space\space\space\space\space\space\space\space \zap}, 7.407957980278479) (\tinyfilemanagershort \texttt{\space\space\space\space\space\space\space\space \zap}, 9.593392630241423) };
\addplot+[ybar] plot coordinates { (\oscommerce \texttt{\space\space\space\space\space \wapiti}, 6.480739523010962) (\oscommerce \texttt{\space\space\space\space\space\space\space\space \zap}, 8.308068500053697) (\piwigo \texttt{\space\space\space\space\space \wapiti}, 9.403921568627451) (\hotcrp \texttt{\space\space\space\space\space \wapiti}, 11.757442721909728) (\hotcrp \texttt{\space\space\space\space \arachni}, 11.67891216142121) (\nextcloud \texttt{\space\space\space\space\space \wapiti}, 19.140723212304962) (\hotcrp \texttt{\space\space\space\space\space\space\space\space \zap}, 18.8301958754461) (\nextcloud \texttt{\space\space\space\space \arachni}, 20.952901386817114) (\oscommerce \texttt{\space\space\space\space \arachni}, 21.946410968724795) (\kanboard \texttt{\space\space\space\space\space \wapiti}, 25.224793190264958) (\piwigo \texttt{\space\space\space\space\space\space\space\space \zap}, 22.92114383070996) (\wordpress \texttt{\space\space\space\space\space \wapiti}, 32.61865739037361) (\dokuwiki \texttt{\space\space\space\space \arachni}, 38.93720322253641) (\kanboard \texttt{\space\space\space\space \arachni}, 38.118723631826015) (\piwigo \texttt{\space \blackostrichshort}, 43.16473460599419) (\piwigo \texttt{\space\space\space\space \arachni}, 44.15056921111344) (\kanboard \texttt{\space \blackostrichshort}, 51.570247933884296) (\hotcrp \texttt{\space \blackostrichshort}, 52.66943517086013) (\kanboard \texttt{\space\space\space\space\space\space\space\space \zap}, 52.273399160409156) (\tinyfilemanagershort \texttt{\space\space\space\space\space \wapiti}, 62.15846994535519) (\wordpress \texttt{\space\space\space\space\space\space\space\space \zap}, 64.87282903942712) (\oscommerce \texttt{\space \blackostrichshort}, 62.74080138701599) (\tinyfilemanagershort \texttt{\space\space\space\space \arachni}, 60.96904441453567) (\nextcloud \texttt{\space\space\space\space\space\space\space\space \zap}, 64.12656352394792) (\tinyfilemanagershort \texttt{\space \blackostrichshort}, 67.12422786547701) (\leantime \texttt{\space\space\space\space \arachni}, 66.61617492096944) (\wordpress \texttt{\space\space\space\space \arachni}, 67.9356749621712) (\leantime \texttt{\space\space\space\space\space \wapiti}, 71.6941331575478) (\dokuwiki \texttt{\space \blackostrichshort}, 83.06912196423805) (\wordpress \texttt{\space \blackostrichshort}, 84.02850589777195) (\dokuwiki \texttt{\space\space\space\space\space \wapiti}, 73.9580764488286) (\nextcloud \texttt{\space \blackostrichshort}, 85.7176003132296) (\leantime \texttt{\space \blackostrichshort}, 82.9102329233185) (\leantime \texttt{\space\space\space\space\space\space\space\space \zap}, 82.73238612348409) (\dokuwiki \texttt{\space\space\space\space\space\space\space\space \zap}, 81.74025073088549) (\tinyfilemanagershort \texttt{\space\space\space\space\space\space\space\space \zap}, 81.70266836086404) };
\addplot+[ybar] plot coordinates { (\oscommerce \texttt{\space\space\space\space\space \wapiti}, 93.51678218271637) (\oscommerce \texttt{\space\space\space\space\space\space\space\space \zap}, 91.68945320567364) (\piwigo \texttt{\space\space\space\space\space \wapiti}, 90.58823529411765) (\hotcrp \texttt{\space\space\space\space\space \wapiti}, 88.19865550830018) (\hotcrp \texttt{\space\space\space\space \arachni}, 88.19223599078846) (\nextcloud \texttt{\space\space\space\space\space \wapiti}, 80.83561339939362) (\hotcrp \texttt{\space\space\space\space\space\space\space\space \zap}, 80.41245538998011) (\nextcloud \texttt{\space\space\space\space \arachni}, 78.8846386743269) (\oscommerce \texttt{\space\space\space\space \arachni}, 76.66783436117606) (\kanboard \texttt{\space\space\space\space\space \wapiti}, 74.69128401870279) (\piwigo \texttt{\space\space\space\space\space\space\space\space \zap}, 75.8107297051364) (\wordpress \texttt{\space\space\space\space\space \wapiti}, 67.27765327624846) (\dokuwiki \texttt{\space\space\space\space \arachni}, 60.75868324174358) (\kanboard \texttt{\space\space\space\space \arachni}, 60.92459445005645) (\piwigo \texttt{\space \blackostrichshort}, 54.87244554664821) (\piwigo \texttt{\space\space\space\space \arachni}, 53.09395320455369) (\kanboard \texttt{\space \blackostrichshort}, 46.824085005903186) (\hotcrp \texttt{\space \blackostrichshort}, 46.20687783296692) (\kanboard \texttt{\space\space\space\space\space\space\space\space \zap}, 46.27801099745758) (\tinyfilemanagershort \texttt{\space\space\space\space\space \wapiti}, 35.040983606557376) (\wordpress \texttt{\space\space\space\space\space\space\space\space \zap}, 33.45954399539057) (\oscommerce \texttt{\space \blackostrichshort}, 34.42095293135555) (\tinyfilemanagershort \texttt{\space\space\space\space \arachni}, 34.79138627187079) (\nextcloud \texttt{\space\space\space\space\space\space\space\space \zap}, 32.176539305353735) (\tinyfilemanagershort \texttt{\space \blackostrichshort}, 30.542210020590254) (\leantime \texttt{\space\space\space\space \arachni}, 29.7418335089568) (\wordpress \texttt{\space\space\space\space \arachni}, 28.217056759280126) (\leantime \texttt{\space\space\space\space\space \wapiti}, 24.752801582069875) (\dokuwiki \texttt{\space \blackostrichshort}, 16.6693354683747) (\wordpress \texttt{\space \blackostrichshort}, 13.828636959370904) (\dokuwiki \texttt{\space\space\space\space\space \wapiti}, 18.204685573366216) (\nextcloud \texttt{\space \blackostrichshort}, 12.310213315158283) (\leantime \texttt{\space \blackostrichshort}, 12.817325307511123) (\leantime \texttt{\space\space\space\space\space\space\space\space \zap}, 12.40693175537276) (\dokuwiki \texttt{\space\space\space\space\space\space\space\space \zap}, 10.851791288836035) (\tinyfilemanagershort \texttt{\space\space\space\space\space\space\space\space \zap}, 8.703939008894537) };

  \end{axis}
  \end{tikzpicture}

  \caption{
The X-axis denotes the application and scanner we are comparing against while
		the Y-axis contains the fraction of unique lines of code the scanners find.
		Starting from the top it shows the fraction only we find, followed by
		the fraction of lines of code found by both scanners and finally
		what only the other scanners find.
      }

    \label{fig:code_coverage}
  \end{figure*}

\newcommand{\yurashort}{Yura\xspace}
\newcommand{\evoshort}{Evo\xspace}
\begin{figure}
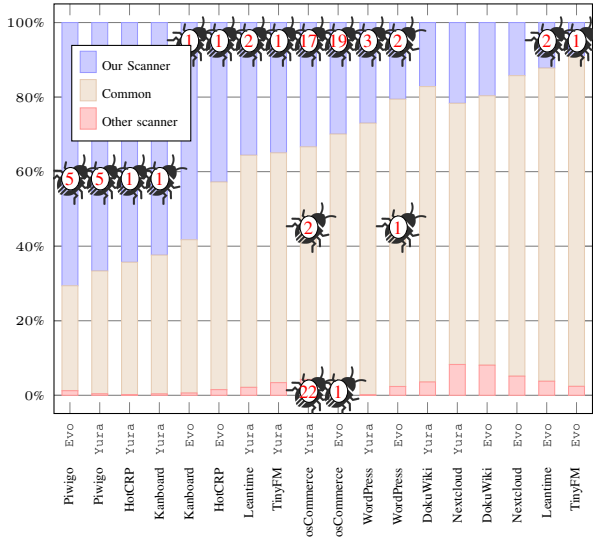

  \begin{tikzpicture}
  \begin{axis}[
      reverse legend,
      ybar stacked,
      bar width=6pt,
      enlargelimits=0.05,
          legend style={font=\small, at={(0.255,0.9)}, cells={align=left}},
          legend cell align={left},
          legend entries={\tiny Other scanner, \tiny Common, \tiny Our Scanner},
      ymajorgrids,
      major grid style={gray},
      enlarge x limits={abs=6pt},
      width=0.48\textwidth,
      height=7cm,
      yticklabel={\tiny\pgfmathprintnumber{\tick}\%},
      cycle list name=mycolorlist,
      symbolic x coords={\piwigo \texttt{\space\space \evoshort},\piwigo \texttt{ \yurashort},\hotcrp \texttt{ \yurashort},\kanboard \texttt{ \yurashort},\kanboard \texttt{\space\space \evoshort},\hotcrp \texttt{\space\space \evoshort},\leantime \texttt{ \yurashort},\tinyfilemanagershort \texttt{ \yurashort},\oscommerce \texttt{ \yurashort},\oscommerce \texttt{\space\space \evoshort},\wordpress \texttt{ \yurashort},\wordpress \texttt{\space\space \evoshort},\dokuwiki \texttt{ \yurashort},\nextcloud \texttt{ \yurashort},\dokuwiki \texttt{\space\space \evoshort},\nextcloud \texttt{\space\space \evoshort},\leantime \texttt{\space\space \evoshort},\tinyfilemanagershort \texttt{\space\space \evoshort}},
      xtick=data,
      x tick label style={rotate=90,anchor=east,font=\tiny},
      ]

\node[text=red] at (0, 580 ) {\scriptsize 5};
\draw[fill=white, opacity=1.0,  xscale=0.1] (2, 580) circle[radius=30];
\node at (1, 580 ) {\includegraphics[width=0.75cm]{bug.png}};

\node[text=red] at (10, 580 ) {\scriptsize 5};
\draw[fill=white, opacity=1.0,  xscale=0.1] (102, 580) circle[radius=30];
\node at (11, 580 ) {\includegraphics[width=0.75cm]{bug.png}};

\node[text=red] at (20, 580 ) {\scriptsize 1};
\draw[fill=white, opacity=1.0,  xscale=0.1] (202, 580) circle[radius=30];
\node at (21, 580 ) {\includegraphics[width=0.75cm]{bug.png}};

\node[text=red] at (30, 580 ) {\scriptsize 1};
\draw[fill=white, opacity=1.0,  xscale=0.1] (302, 580) circle[radius=30];
\node at (31, 580 ) {\includegraphics[width=0.75cm]{bug.png}};

\node[text=red] at (40, 950 ) {\scriptsize 1};
\draw[fill=white, opacity=1.0,  xscale=0.1] (402, 950) circle[radius=30];
\node at (41, 950 ) {\includegraphics[width=0.75cm]{bug.png}};

\node[text=red] at (50, 950 ) {\scriptsize 1};
\draw[fill=white, opacity=1.0,  xscale=0.1] (502, 950) circle[radius=30];
\node at (51, 950 ) {\includegraphics[width=0.75cm]{bug.png}};

\node[text=red] at (60, 950 ) {\scriptsize 2};
\draw[fill=white, opacity=1.0,  xscale=0.1] (602, 950) circle[radius=30];
\node at (61, 950 ) {\includegraphics[width=0.75cm]{bug.png}};

\node[text=red] at (70, 950 ) {\scriptsize 1};
\draw[fill=white, opacity=1.0,  xscale=0.1] (702, 950) circle[radius=30];
\node at (71, 950 ) {\includegraphics[width=0.75cm]{bug.png}};

\node[text=red] at (80, 950 ) {\scriptsize 17};
\draw[fill=white, opacity=1.0,  xscale=0.1] (802, 950) circle[radius=30];
\node at (81, 950 ) {\includegraphics[width=0.75cm]{bug.png}};

\node[text=red] at (80, 450 ) {\scriptsize 2};
\draw[fill=white, opacity=1.0,  xscale=0.1] (802, 450) circle[radius=30];
\node at (81, 450 ) {\includegraphics[width=0.75cm]{bug.png}};

\node[text=red] at (80, 10 ) {\scriptsize 22};
\draw[fill=white, opacity=1.0,  xscale=0.1] (802, 10) circle[radius=30];
\node at (81, 10 ) {\includegraphics[width=0.75cm]{bug.png}};

\node[text=red] at (90, 950 ) {\scriptsize 19};
\draw[fill=white, opacity=1.0,  xscale=0.1] (902, 950) circle[radius=30];
\node at (91, 950 ) {\includegraphics[width=0.75cm]{bug.png}};

\node[text=red] at (90, 10 ) {\scriptsize 1};
\draw[fill=white, opacity=1.0,  xscale=0.1] (902, 10) circle[radius=30];
\node at (91, 10 ) {\includegraphics[width=0.75cm]{bug.png}};

\node[text=red] at (100, 950 ) {\scriptsize 3};
\draw[fill=white, opacity=1.0,  xscale=0.1] (1002, 950) circle[radius=30];
\node at (101, 950 ) {\includegraphics[width=0.75cm]{bug.png}};

\node[text=red] at (110, 950 ) {\scriptsize 2};
\draw[fill=white, opacity=1.0,  xscale=0.1] (1102, 950) circle[radius=30];
\node at (111, 950 ) {\includegraphics[width=0.75cm]{bug.png}};

\node[text=red] at (110, 450 ) {\scriptsize 1};
\draw[fill=white, opacity=1.0,  xscale=0.1] (1102, 450) circle[radius=30];
\node at (111, 450 ) {\includegraphics[width=0.75cm]{bug.png}};

\node[text=red] at (160, 950 ) {\scriptsize 2};
\draw[fill=white, opacity=1.0,  xscale=0.1] (1602, 950) circle[radius=30];
\node at (161, 950 ) {\includegraphics[width=0.75cm]{bug.png}};

\node[text=red] at (170, 950 ) {\scriptsize 1};
\draw[fill=white, opacity=1.0,  xscale=0.1] (1702, 950) circle[radius=30];
\node at (171, 950 ) {\includegraphics[width=0.75cm]{bug.png}};

\addplot+[ybar] plot coordinates { (\piwigo \texttt{\space\space \evoshort}, 1.2356199403493822) (\piwigo \texttt{ \yurashort}, 0.3984375) (\hotcrp \texttt{ \yurashort}, 0.19178607633085837) (\kanboard \texttt{ \yurashort}, 0.37654653039268426) (\kanboard \texttt{\space\space \evoshort}, 0.6082289803220036) (\hotcrp \texttt{\space\space \evoshort}, 1.51928847557514) (\leantime \texttt{ \yurashort}, 2.130506036991204) (\tinyfilemanagershort \texttt{ \yurashort}, 3.3944331296673456) (\oscommerce \texttt{ \yurashort}, 4.2538718301615175) (\oscommerce \texttt{\space\space \evoshort}, 2.402683265069178) (\wordpress \texttt{ \yurashort}, 0.10703057060672955) (\wordpress \texttt{\space\space \evoshort}, 2.3588067020841845) (\dokuwiki \texttt{ \yurashort}, 3.571059964908659) (\nextcloud \texttt{ \yurashort}, 8.263217891651287) (\dokuwiki \texttt{\space\space \evoshort}, 8.077528532073986) (\nextcloud \texttt{\space\space \evoshort}, 5.130868009262508) (\leantime \texttt{\space\space \evoshort}, 3.7687450670876084) (\tinyfilemanagershort \texttt{\space\space \evoshort}, 2.400548696844993) };
\addplot+[ybar] plot coordinates { (\piwigo \texttt{\space\space \evoshort}, 28.150056164542743) (\piwigo \texttt{ \yurashort}, 32.951171875) (\hotcrp \texttt{ \yurashort}, 35.494123126661) (\kanboard \texttt{ \yurashort}, 37.23029107644492) (\kanboard \texttt{\space\space \evoshort}, 41.115086463923674) (\hotcrp \texttt{\space\space \evoshort}, 55.705441864237244) (\leantime \texttt{ \yurashort}, 62.26964112512124) (\tinyfilemanagershort \texttt{ \yurashort}, 61.642905634758996) (\oscommerce \texttt{ \yurashort}, 62.40013921187098) (\oscommerce \texttt{\space\space \evoshort}, 67.68149127616344) (\wordpress \texttt{ \yurashort}, 72.89785269917721) (\wordpress \texttt{\space\space \evoshort}, 77.07069881487536) (\dokuwiki \texttt{ \yurashort}, 79.2445040767881) (\nextcloud \texttt{ \yurashort}, 70.06296819020736) (\dokuwiki \texttt{\space\space \evoshort}, 72.2697756788666) (\nextcloud \texttt{\space\space \evoshort}, 80.62030734685285) (\leantime \texttt{\space\space \evoshort}, 84.01407524335701) (\tinyfilemanagershort \texttt{\space\space \evoshort}, 88.47736625514403) };
\addplot+[ybar] plot coordinates { (\piwigo \texttt{\space\space \evoshort}, 70.61432389510787) (\piwigo \texttt{ \yurashort}, 66.650390625) (\hotcrp \texttt{ \yurashort}, 64.31409079700813) (\kanboard \texttt{ \yurashort}, 62.39316239316239) (\kanboard \texttt{\space\space \evoshort}, 58.27668455575432) (\hotcrp \texttt{\space\space \evoshort}, 42.77526966018761) (\leantime \texttt{ \yurashort}, 35.59985283788755) (\tinyfilemanagershort \texttt{ \yurashort}, 34.96266123557366) (\oscommerce \texttt{ \yurashort}, 33.345988957967506) (\oscommerce \texttt{\space\space \evoshort}, 29.915825458767376) (\wordpress \texttt{ \yurashort}, 26.99511673021607) (\wordpress \texttt{\space\space \evoshort}, 20.57049448304046) (\dokuwiki \texttt{ \yurashort}, 17.18443595830323) (\nextcloud \texttt{ \yurashort}, 21.673813918141352) (\dokuwiki \texttt{\space\space \evoshort}, 19.652695789059425) (\nextcloud \texttt{\space\space \evoshort}, 14.248824643884639) (\leantime \texttt{\space\space \evoshort}, 12.21717968955538) (\tinyfilemanagershort \texttt{\space\space \evoshort}, 9.122085048010973) };

  \end{axis}
  \end{tikzpicture}

  \caption{Unique LoC compared between \shortname and the latest SOTA crawlers following the format of \Cref{fig:code_coverage}.}

    \label{fig:code_coverage_recent_sota}
  \end{figure}

\subsubsection{Vulnerabilities}
\label{sec:eval_vulns}
Here we present the XSS vulnerabilities found by each scanner.
We report the \textit{verified} vulnerabilities, which we have manually checked.
Verifying the vulnerabilities is important, as scanners can mistakenly report safe parts of the application as vulnerable.
Across almost all applications, our method finds more XSS vulnerabilities than the other scanners.
In total, we find \totalXSS verified XSS vulnerabilities across all applications, compared to \othersXSS for the other scanners (\blackostrich and \evocrawl find one duplicate on \wordpress).
We see that the compared state-of-the-art scanners do not find any XSS vulnerabilities in many of these modern applications.
Excluding the 24 vulnerabilities \yurascanner finds in \oscommerce, all other scanners only find 5 XSS in \othersXSSApplications applications.
Our method also finds most of these vulnerabilities.

An exception is \arachni, which finds two more vulnerabilities in \tinyfilemanager.
It should be noted that \arachni has a high rate of false positives in this case,
because of its imprecise method that reuses a shared payload.
After finding one stored vulnerability, all parameters tested afterwards will appear vulnerable.
During our manual verification of its 10 reported vulnerabilities, 3 of these were confirmed to be valid.
However,  we could not determine if \arachni actually found them or happened to mark all parameters
after the first successful stored injection.
All other scanners consistently use unique payloads, and avoid any false positives.

We note that while the other scanners cover thousands of LoC that we miss (see \Cref{sec:eval_coverage}), they do not find as many vulnerabilities in this code.
This might indicate that scanners should focus more on \textit{intended} or high-quality code coverage, which we discuss in \Cref{sec:analysis_intended}.

\begin{table}[h!]
	\caption{ This table presents the \textit{verified} XSS injections from the scanners.
	We discuss \arachni's FPs (*) in \Cref{sec:eval_vulns}.
	\wapiti and \zap are excluded as they report no XSS.}
	\label{tab:injections}
	\begin{center}
		\begin{adjustbox}{width=0.48\textwidth}
		\begin{tabular}{  l | c | c | c | c | c}
			\toprule
			Scanner & \arachni & \blackostrich & \shortname & \yura & \evocrawl \\
			\midrule
			\dokuwiki               & 0              & 0   	& 0     & 0    & 0 \\
			\hotcrp                 & 0              & 0   	& 1     & 0    & 0  \\
			\kanboard               & 0              & 0   	& 1     & 0    & 0 \\
			\leantime               & 0              & 0   	& 2     & 0    & 0  \\
			\nextcloud              & 0              & 0   	& 0     & 0    & 0 \\
			\oscommerce             & 0              & 0   	& 19    & 24   & 1 \\
			\piwigo                 & 0              & 0   	& 5     & 0    & 0 \\
			\tinyfilemanager        & \enskip 3$^*$  & 0    & 1     & 0    & 0  \\
			\wordpress              & 0              & 1   	& 3     & 0    & 1 \\

			\bottomrule
		\end{tabular}
		      \end{adjustbox}
\end{center}
\end{table}

\subsection{Recent State-Of-The-Art}
\label{sec:recent_sota}
We separately evaluate state-of-the-art scanners \evocrawl and \yura, comparing coverage and vulnerabilities in \Cref{fig:code_coverage_recent_sota}.
We improve coverage by \improvedCoverageVSSota compared to the union of \evocrawl and \yura.
We also find more unique LoC across all evaluated applications.
Still, both \evocrawl and \yurascanner find thousands of LoC that we miss.
Notably, in deeper applications like \hotcrp (with functionality for reviewing papers dependent on submitting a paper first), we perform better than \yurascanner.
This can be due to \yurascanner's task-solving missing dependent functionality (see \Cref{appendix:task_solvers}).
We note that the novel element detection of \shortname leads to its much improved performance on \piwigo.

\shortname finds the most XSS across the most applications, showcasing a general improvement across
these diverse applications.
While \yura performs better than \shortname on \oscommerce, this is the only application where they find any XSS.
This could be due to a biased set of evaluated applications,
but the results  are mirrored in \yura's own evaluated  applications~\cite{DBLP:conf/ndss/StafeevRSKP25}.
Moreover, while \evocrawl has high coverage on \oscommerce, it only finds one vulnerability. %
This highlights the importance of interacting in a security-relevant manner.

\subsection{Repeated Evaluation Case Study}
\label{sec:repeated_evaluation}
We perform a limited case study of repeated runs for both \shortname and \yura.
While a single extra run does not provide statistical significance it
can still provide valuable insights.

\para{Coverage}
In \Cref{table:repeated_evaluation} we present the coverage achieved by both runs for \shortname and \yura.
Most results are relatively close, with a difference of about 10\% compared to the max.
However, there are some notable exceptions, including \piwigo where \shortname manages to
destroy the state by attempting to update the entire web application, and breaking it.
We note that \shortname is the only scanner that reaches this point in the
application and discuss it in \Cref{sec:destroying_state}.

By comparing the union of the two runs to the max we can learn more about how \textit{saturated} the
scanner is on a particular application---are the same parts of the application being explored?
For example, on \tinyfilemanager the same code is exercised, while
on \oscommerce both scanners uncover thousands of new lines of code between the runs.
This indicates that running the scanner multiple times or for longer periods on \tinyfilemanager
will have less of an effect than doing the same on \oscommerce.

\begin{table}
\centering
\caption{Coverage results for the repeated evaluation of \shortname and \yura. We present
	the max and min coverage achieved, as well as the union of both runs.}
\label{table:repeated_evaluation}

\begin{adjustbox}{max width=0.48\textwidth}
\begin{tabular}{l | rrr | rrr }
	\toprule
		Scanner     & \multicolumn{3}{c|}{\shortname}  & \multicolumn{3}{c}{\yura}  \\
		Application & Max & Min & Union   & Max & Min & Union  \\
            \midrule
	    \dokuwiki          & \numprint{18686}   & \numprint{18632}   & \numprint{19342}   & \numprint{18249}   & \numprint{16048}   & \numprint{18354}   \\
		\hotcrp            & \numprint{36429}   & \numprint{36125}   & \numprint{37848}   & \numprint{15233}   & \numprint{13025}   & \numprint{16356}   \\
		\kanboard          & \numprint{16770}   & \numprint{16668}   & \numprint{17060}   & \numprint{6292}    & \numprint{4263}    & \numprint{6494}    \\
		\leantime          & \numprint{29918}   & \numprint{29262}   & \numprint{30592}   & \numprint{21715}   & \numprint{19255}   & \numprint{22357}   \\
		\nextcloud         & \numprint{67599}   & \numprint{64854}   & \numprint{68623}   & \numprint{60335}   & \numprint{57717}   & \numprint{64005}   \\
		\oscommerce        & \numprint{126361}  & \numprint{121048}  & \numprint{132901}  & \numprint{91575}   & \numprint{84268}   & \numprint{99675}   \\
		\piwigo            & \numprint{50996}   & \numprint{28121}   & \numprint{52808}   & \numprint{17075}   & \numprint{15765}   & \numprint{18161}   \\
		\tinyfilemanager   & \numprint{1446}    & \numprint{1423}    & \numprint{1454}    & \numprint{1064}    & \numprint{958}     & \numprint{1107}    \\
		\wordpress         & \numprint{59732}   & \numprint{58937}   & \numprint{62017}   & \numprint{43654}   & \numprint{28625}   & \numprint{45723}   \\

\bottomrule
\end{tabular}
\end{adjustbox}
\end{table}

\para{Vulnerabilities}
The vulnerabilities found in the repeated evaluation are mostly the same.
\yurascanner finds \yuraRepeatXSS vulnerabilities, all in \oscommerce,
with \yuraRepeatOverlapYuraXSS XSS also found by \yurascanner in the main evaluation.
On \oscommerce, \shortname finds \ourRepeatOscommerceXSS vulnerabilities, compared to \ourOscommerceXSS
in the main evaluation, with an overlap of \ourOscommerceOverlapXSS.
This indicates that while both scanners are relatively stable on \oscommerce,
running each scanner for longer periods or multiple times could reduce false negatives, as we discuss in \Cref{sec:discussion_time}.

\shortname finds \ourRepeatXSS XSS across \ourRepeatXSSApplications applications
in the repeated run, missing a \kanboard vulnerability.
While \shortname is able to find and submit the
correct vulnerable form, it does not resubmit it with a payload.
We suspect this was because of a timeout in the retracing.
On \piwigo, \shortname only finds 2 XSS in this run, most likely due to
the state-destroying action of updating the application.
However, this run performed better
on \leantime, finding 9 XSS as opposed to 2 in the initial run.
All 7 new XSS come from the same XSS-vulnerable form.
Upon inspection, the first run misses this due to randomness in navigation.
In the first run, \shortname updates the current
username, then the LLM detects the form as ``dangerous'' and skips it.
In the second run, the form is filled out first, and the LLM does not see it as dangerous.

\subsection{State Depth Case Study}
\label{sec:state_depth}

We not only aim to increase coverage and vulnerabilities found, but to achieve this by reaching deeper states in the application, a goal shared with \yurascanner.
Therefore, we perform a limited case study on the states of two smaller applications, and compare against the latest \evocrawl and \yurascanner.
However, as state depth is not well-defined for scanners, we first define a metric.
\yurascanner uses trace length, in terms of steps/actions, as a proxy for state depth.
However, there are some natural properties for state depth that should be satisfied, which trace length does not:

\para{Properties of state depth}
State depth ($D$) should consider the application's state before any action is taken.
If, for example, the scanner is already logged in, logging in another time does not increase $D$.
State depth should be consistent across a given trace.
If a set of actions is repeated, such as logging in and out, this should not increase $D$.
Finally, state depth should be consistent across all traces for the application.
$D$ should measure the shortest path length, in terms of actions, to particular states.
Ideally, the shortest path would be calculated from all possible interactions with the web application.
When working with the real traces, state depth can at least be consistent among all of the scanner traces.

\para{Event traces}
We define events as the execution of source files (e.g., \texttt{index.php}) or lines within them (e.g., \texttt{(index.php, 6)}).
Alternatively, an annotated set can be provided by a developer or security analyst, describing which of these source files or lines is relevant.
From coverage logs, %
we observe how each server-side request made by the scanner results in the execution of a set of events.
The first execution of each of these events forms a trace of the states a given scanning run reaches. %

\hotcrp's developers annotate which source files correspond to `pages' and their parts by a leading comment matching the regex: \lstinline{^// .*\.php -- HotCRP (.+) page$}.
\tinyfilemanager requires more granular annotations, due to its implementation in a single PHP source file.
We generated LoC-level state event annotations with manual code analysis, and the results are included in \Cref{appendix:tinyfilemanager}.

\para{State graphs}
State graphs are inferred using the evidence from traces about relationships between events.
Each event forms a node in the state graph.
We make these graphs in three steps.

First, we make a happens-before graph that relates each event in these traces.
If, across all traces, event A always happens before B, then edge A\textrightarrow B is formed.

Second, edge A\textrightarrow B is only kept if there are no common events that always occur between A and B.
A\textrightarrow B will be kept if the set of traces only shows different events in between the two: [[A,C,B], [A,D,B]], or even [[A,B], [A,C,B]].
Otherwise, A\textrightarrow B will be removed if there is always event C between them: [[A,C,B], [A,C,D,B]].

Third, the graph can be simplified:
If an event always happens with another event, these nodes can be merged.
Upward edges from nodes of greater depth to nodes of lesser depth can be pruned.
The resulting state graph contains the shortest paths to relevant events for particular application states.

\para{State depth}
In \Cref{table:state_depths} we summarize the deepest states each scanner achieves in \hotcrp and \tinyfilemanager.
We also report the average terminating depth and the number of states found by each scanner.
Average terminating depth is defined as the mean length of paths to the deepest, terminating states found by that scanner.
This excludes paths to intermediary states, to quantify how deep each scanner gets in application functionality.
Notably, in \hotcrp \shortname discovers all states that any scanner finds.

\begin{table}
\centering
\caption{Scanner state depths on \hotcrp and \tinyfilemanager. The max depth ($\max(D)$), average terminating depth ($\bar{D}$), and total number of states ($S$) are reported.
}
\label{table:state_depths}
\begin{adjustbox}{max width=0.48\textwidth}
\begin{tabular}{l | rrr | rrr}
	\toprule
	\multirow{2}{*}{Scanner} & \multicolumn{3}{c|}{\hotcrp} & \multicolumn{3}{c}{\tinyfilemanager} \\
	& $\max(D)$ & $\bar{D}$ & $S$ & $\max(D)$ & $\bar{D}$ & $S$ \\
	\midrule
\arachni & 2 & 1.38 & 13 & 4 & 3.05 & 75 \\
\blackostrich & 4 & 2.80 & 100 & 3 & 3.00 & 52 \\
\evocrawl & 3 & 2.90 & 80 & 4 & 3.05 & 75 \\
\rowcolor{lightgray} \shortname & 4 & 3.14 & 121 & 4 & 3.11 & 97 \\
\wapiti & 2 & 1.38 & 13 & 4 & 2.95 & 74 \\
\yurascanner & 3 & 2.21 & 53 & 3 & 2.93 & 39 \\
\zap & 2 & 1.75 & 23 & 5 & 3.61 & 101 \\
	\midrule
All scanners & 4 & 3.14 & 121 & 5 & 3.61 & 119 \\
	\bottomrule
\end{tabular}
\end{adjustbox}
\end{table}

\para{\hotcrp states}
\label{sec:hotcrp_states}
In \Cref{fig:hotcrp_states} we visualize the state graph of \hotcrp using the earlier page definition, and compare \shortname with the latest \evocrawl and \yurascanner.
In this graph, each scanner starts by requesting the home page (\texttt{index.php}).
Requests are funneled by \hotcrp through code that broadly applies across pages (e.g., delegation, JSON API access).
At depth 2, we have states available in a single action after the home page.
These include settings, help, or starting a paper submission.
All three compared scanners reach these states, except \evocrawl missing the help page.
After these nodes, we start seeing deep application functionality at depth 3.
In particular, the graph's right-hand side shows the states unlocked after a paper is submitted, including various ways to assign reviews for this submission.
\yurascanner misses many of these nodes, as it never completes a paper submission.
Finally, at depth 4, there are only states reached by \shortname.
Most notably, we have the actual paper review page, and drawing the review preference graph.

\begin{figure*}[tb]
	\centering
	\includegraphics[width=\textwidth]{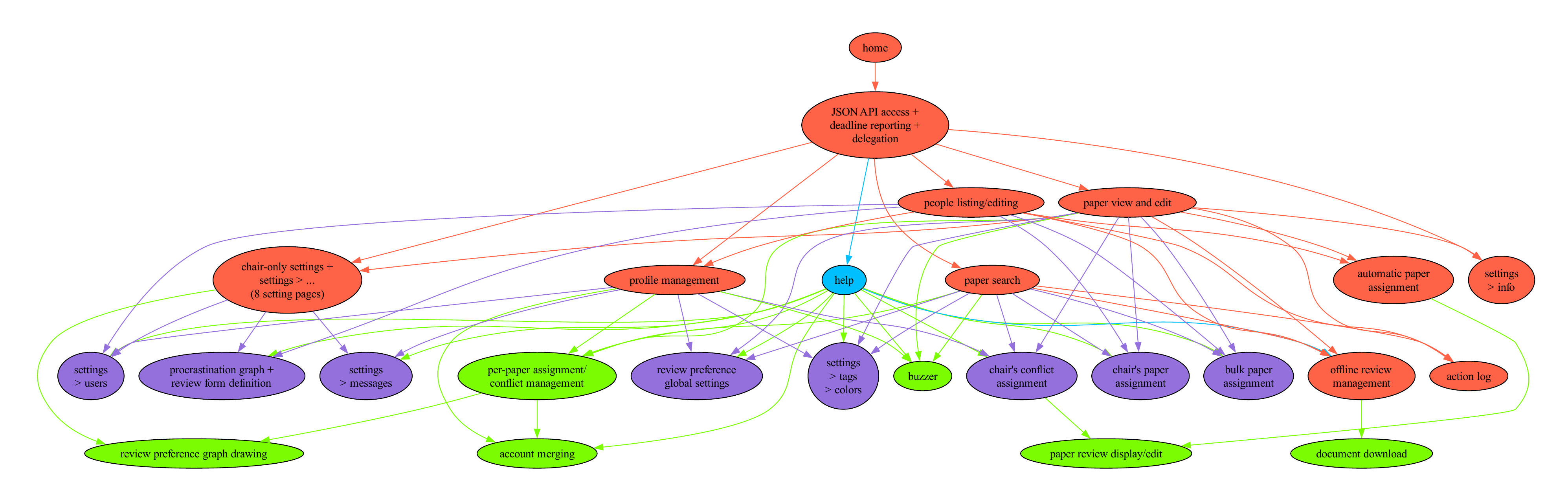}
	\caption{
		In \hotcrp, \shortname finds all pages in this state graph of pages.
		Both \evocrawl and \yurascanner find red.
		\evocrawl can also find purple, while \yurascanner can also find blue.
		Green are only found by \shortname.}
	\label{fig:hotcrp_states}
\end{figure*}

\para{\tinyfilemanager states}
\label{sec:tinyfilemanager_states}
In \Cref{appendix:tinyfilemanager} we visualize the state graph of \tinyfilemanager using our manual annotations of relevant LoC.
\yurascanner again performs worst, \evocrawl improves substantially, and \shortname does slightly better than both.
Surprisingly, \zap reaches states that other scanners miss.
In this case, \zap's simple heuristics succeed due to their tendency to avoid unnecessary fields and accept default values.
Due to the simplicity of this application, our manual annotations include \emph{all} relevant states in the application.
Therefore, we can measure how scanners are performing against \emph{ground truth}.
There is substantial room for improvement: only 16/39 of these states are reached by any scanner.

\para{Limitations}
The core limitation of our state depth definition is in the quality of the set of traces to generate the graph.
If scanners fail to explore different paths through the application, and do not overlap in what states they discover, the graph will overestimate which prior states are necessary for later ones.
We attempt to increase the quality of our set of traces by adding more scanner runs, which are otherwise not included in the evaluation.
Future work could also integrate traces from test suites or users.

\subsection{Client-side Coverage}
\label{sec:clientside_coverage}

As modern web applications rely heavily on JavaScript we make a first attempt to
report on the \textit{client-side coverage} achieved by our scanner.
This goes beyond the efforts of previous works that focused server-side coverage,
or coarse-grained network fetches of JS scripts~\cite{DBLP:conf/uss/StafeevP24}.

Using the built-in profiler in Chrome's DevTools~\cite{devtoolsprofiler} we collect byte-level coverage metrics in our scanner.
As the profiler resets when a document is unloaded, we collect coverage  both before and after each action.
Our analysis is limited to scripts that have a URL to ensure the code comes from the web application,
as opposed to scripts injected by the scanner. This entails that we lack coverage data for anonymous functions and code inside \texttt{eval}s.
Furthermore, we deduplicate scripts by their content hash. Therefore, the same script loaded on different pages will be counted as the same script.

To better understand how our coverage compares with another scanner, we modify \blackostrich to collect client-side coverage.
Note that in general it is not trivial to add this coverage collection as it requires knowledge of where in the scanner's code actions are performed.
Additionally, it also requires the scanner to use Chrome.
As \blackostrich uses both Chrome and a similar crawling loop to our method,  it is suitable for a comparison.

In \Cref{table:clientside_compare_blackostrich} we present a comparison between \shortname and \blackostrich. The table contains both the byte-level coverage achieved  by the scanners and total bytes from all loaded scripts, as reported by the Chrome profiler.
As is evident, \shortname performs better across all applications.
\shortname's coverage improvements range from \clientCoverageImprovedkanboard on \kanboard to \clientCoverageImprovedpiwigo on \piwigo, with an average of \clientCoverageImprovedAvg.
While we cover a majority of the bytes (\clientCoveragePercent), there is still room for future improvement.
At the same time, it is important to note that the total number of loaded bytes can contain dead or unreachable code.

\subsection{Non-semantic Interactable Elements}
\label{sec:nonsemantic_interactable}

Based on the scanner logs from \shortname we calculate the number of unique interactable elements we find
and, of those, the number of non-semantic ones.
While no definitive list of \textit{semantically interactable tags} exists, we based ours on the ones defined in \evocrawl~\footnote{ Elements of interest in \evocrawl: \texttt{input, button, textarea, select, a}. }.
In \Cref{table:nonsemantic_interactable} we present both the number of elements we find and the number of non-semantic tags per application.
Some applications have notably more non-semantic elements and tags (e.g., \nextcloud and \oscommerce).
Across all applications we find thousands of non-semantic elements that
scanners using a purely semantic selection method would miss.
We note that \verb|<li>| tags are popular for menus while \verb|<span>| and \verb|<label>|
are common for small clickable icons, for example a plus sign for adding.

\subsection{Ablation Study Results}
\label{sec:ablation_study_summary}

We perform two ablation studies to determine the effects of the LLM-powered form solving  and element discovery methods used in \shortname.
\ablatedLLM removes LLM form solving, while \ablatedELM removes the improved interactable element discovery method.
We describe the setup and results in more detail in \Cref{sec:ablation_study}, and summarize the main findings here.

\para{Coverage}
On average, adding the LLM or element discovery method improves coverage by \improvedCoverageVSbwinputs and \improvedCoverageVSbwelements respectively.
For applications that require solving complex forms early, such as \kanboard, \hotcrp, and \leantime,
the LLM method shows a clear benefit.
For applications like \oscommerce there is a smaller difference in coverage, which we discuss in \Cref{sec:oscommerce_coverage}.
The \ablatedELM version also performs worse
because the scanner cannot
accurately detect elements.
\piwigo's \texttt{label} tag buttons are not found by \ablatedELM, leading to a decrease in coverage, even compared to \ablatedLLM.

\para{Vulnerabilities}
The ablated \ablatedLLM finds \ablatedLLMTotalXSS XSS across \ablatedLLMXSSApplications applications,
while
the \ablatedELM ablation finds \ablatedELMTotalXSS XSS across \ablatedELMXSSApplications applications.
We note a large overlap of \ablatedLLMOverlapXSS vulnerabilities between \shortname and the ablated \ablatedLLM, and \ablatedELMOverlapXSS with \ablatedELM.
Neither ablated version finds XSS in as many applications as \shortname.
While both ablations find many vulnerabilities in \oscommerce, their XSS count is further inflated
by a single form in \leantime with 7 XSS vulnerabilities.
Adding more time or repeated runs is helpful to minimize false negatives as discussed in \Cref{sec:false_negatives}, and these are discovered by \shortname in \Cref{sec:repeated_evaluation}

\section{Analysis / Discussion} %
\label{sec:analysis}

\subsection{The Future of Black-box Crawling}
\label{sec:future_work}

The latest generation of crawlers has focused on improving their crawling strategy with machine learning~\cite{DBLP:conf/ndss/GuoKLL25, DBLP:conf/dsn/CazzaroCKSP25} and sometimes LLMs specifically~\cite{DBLP:journals/corr/abs-2504-20801, DBLP:conf/ndss/StafeevRSKP25}.
Our method and results are complementary with these approaches.
LLM form solving and interactable element detection are two building blocks that can improve these and future work.
Based on our results,
form solving with LLMs should replace prior heuristics,
and our novel interactable element detection provides more accurate identification than prior methods.
At the same time, our relatively simple crawling strategy could be improved by these approaches' adaptability to workflows.

However, researchers must carefully evaluate the performance of black-box LLMs at crawling.
Our results show that LLMs, including smaller models, are already adept at the task of form solving.
At the same time, our evaluation shows that a fully-LLM approach in \yurascanner performs worse overall.
This could be due to the single-pass setup of their task solver (see \Cref{appendix:task_solvers}), but also could be due to the abstract page representation the LLM operates on.
The interactable element detection we propose can improve future page abstractions by accurately identifying actions on the page.

\subsection{Exploring Intended Code Paths}
\label{sec:analysis_intended}

One goal of our approach was to improve the exploration of
what can be seen as ``intended code paths''---paths in the application that the developer expects a user to follow.
Note that this is opposed to the classic goal of ``fuzzing'' the application with unintended values to find error-related code paths.

We believe that current web scanners focus too much on %
these unintended paths, and thus they fail to explore much of the application's deeper functionality (and, in turn, miss vulnerabilities).
Therefore, simply using total server-side code coverage as a metric for performance can be misleading, because it includes non-security-relevant code in unintended paths.
For example, \Cref{sec:eval_coverage} shows that \zap has similar total coverage on \tinyfilemanager compared to our approach.
However, part of \zap's unique coverage is due to discovering error-handling code relating to incorrect CSRF tokens, because \zap fuzzes all request parameters on the HTTP level.
In contrast, our method intentionally submits the form successfully with valid tokens, avoiding some unintended paths.
\shortname still incorporates randomness and injects XSS payloads, triggering unexpected behaviors and error states.

We cannot claim either method of exploration is strictly superior.
If anything, our results suggest that the research community should explore a combined approach or use of multiple scanners to ensure thorough security testing of modern web applications.
As shown in \Cref{sec:tinyfilemanager_states}, much of the functionality in web applications is still unexplored.

\subsection{State Destruction}
\label{sec:destroying_state}
A surprising problem we faced as scanners improve is their increased potential to break the application.
\shortname performs relatively poorly on \piwigo in the repeated evaluation (see \Cref{sec:repeated_evaluation}).
This is because our method is the only system that finds an update mechanism that breaks the application, creating an irrecoverable program state.
A similar problem has also been encountered by \yurascanner, when they delete their user in the web application~\cite{DBLP:conf/ndss/StafeevRSKP25}.
While our crawler can sometimes avoid state-destroying actions (see \Cref{llm-crawl-subprompts}), such as in \dokuwiki where the scanner was prevented from removing the current user,
future research could try to improve avoiding dangerous actions.
Future methods could use an LLM approach similar to ours or
by allowing the scanner to reset the application
such as Enemy of the State~\cite{DBLP:conf/uss/DoupeCKV12},
using heuristics to detect breakage~\cite{DBLP:conf/uss/0001EDS24}.

\subsection{\oscommerce Coverage}
\label{sec:oscommerce_coverage}
While our method generally improves coverage, we still
miss thousands of lines of code on \oscommerce,
that both the latest SOTA, \evocrawl and \yurascanner (\Cref{table:locs_sota}), and our ablated versions, \ablatedLLM and \ablatedELM, find.
This performance difference can be attributed to \oscommerce having a large, primarily shallow and broad surface.
Multiple scans tend to cover different parts of this application.
This favors BFS-like strategies in the limited evaluation time, rather than our navigation strategy which is closer to a DFS for deep crawling.
Still, even if a scanner has a DFS-like strategy, if the scanner cannot solve forms that limit deeper navigation, such as \ablatedLLM,
the resulting behavior tends toward BFS.
Similarly, if the scanner cannot detect the interactable elements necessary for deeper navigation, it also tends toward BFS.

\subsection{False Negatives and Length of Scan}
\label{sec:false_negatives}
\label{sec:discussion_time}
During development \shortname found vulnerabilities
that were not reproduced in the evaluation.
Similarly, the multiple runs compared to the ablations and the latest state-of-the-art \evocrawl and \yurascanner find different sets of vulnerabilities.
We still report these to the developers for ethical reasons.
Our scanner still supports each interaction necessary to find these additional vulnerabilities.
We believe that the final scan did not find them due to two practical limits:
(1) evaluation time limit of 8 hours, and
(2) we destroy the application state (\Cref{sec:destroying_state}).
There is no consensus on the evaluation time for web scanning, with times ranging from 4 hours~\cite{DBLP:conf/ndss/StafeevRSKP25} to 24 hours~\cite{DBLP:conf/ndss/GuoKLL25}.
Interestingly, academic scanners
show potential to continue increasing coverage after 8 hours.
With the scanner's additional randomness,
a longer evaluation could lead to increased code coverage.
Future studies should ideally incorporate both multiple runs and longer evaluation times.

\nextcloud and \dokuwiki are noteworthy for having zero vulnerabilities found across all scanners.
\nextcloud, to our knowledge, does not have any false negatives.
However, manual analysis of \dokuwiki reveals that a relatively simple XSS exists
in the configuration form.
This form contains multiple server-side validations that make it difficult
to solve the entire form. For example, while \shortname provides the
correct type of data, e.g., a path for a folder, the server also
checks that this folder exists. %
The entire form becomes increasingly difficult to successfully submit when needing to combine valid values
with XSS payloads.
However, a \textit{partial form attack} where only one element is
updated, e.g., the title, would successfully submit the form with a payload.
This could be an interesting avenue for future scanners to explore.

\subsection{Exploitability of Vulnerabilities}
\label{sec:exploitability}
Scanners generally do not consider exploitability, e.g.,
a payload resulting in JavaScript execution but is also protected by CSP, CSRF, \texttt{samesite} cookies, or only allowed by administrators (self-XSS).
As scanners use valid admin accounts and CSRF tokens they are not affected by
these countermeasures as a real attacker would be.

We define a vulnerability as \textit{exploitable} if it allows a user to expand their set of permissions to gain additional privileges.
In total, we find that \exploitableXSS out of the \totalXSS XSS vulnerabilities are exploitable.
\kanboard is protected by CSP but they have previously fixed XSS regardless~\cite{kanboardxss}.
In \piwigo, permissions prevent one of the vulnerabilities from being exploitable because only the superuser (webmaster) can change the gallery title.
Three reflected XSS are protected by CSRF, but the last one is stored and does not require webmaster privileges.

\para{Disclosure}
We have disclosed all vulnerabilities and we are happy to report that Kanboard has already fixed the vulnerability we found and TinyFileManager is working on a fix.
The WordPress case is more complex, as their developers consider both the \textit{administrator} and \textit{editor} roles to be trusted and thus won't fix it.
We are still awaiting responses from \leantime, \oscommerce, and \piwigo.

\section{Related Work}
\label{sec:related_work}

\para{Static Analysis}
Static analysis of application source code has been applied~\cite{DBLP:conf/uss/DahseH14, fang2019tap, DBLP:conf/dsn/HuangLZD19} to detect various
vulnerabilities.
Dahse et al.~\cite{DBLP:conf/uss/DahseH14} analyze PHP to find stateful second-order vulnerabilities by examining dataflows.
Huang et al.~\cite{DBLP:conf/dsn/HuangLZD19} find file upload vulnerabilities in PHP with symbolic analysis.
Fang et al.~\cite{fang2019tap}
adapt dataflow analysis to use deep learning.
Restler~\cite{DBLP:conf/icse/AtlidakisGP19} analyzes only a REST API specification. %
ReactAppScan~\cite{DBLP:conf/ccs/GuoKVGC24} targets React single-page applications with abstract interpretation.
With our general approach, we can also detect and interact with React elements.
While the applications in the evaluation mostly do not use React, we can %
add and mark notes as completed in the TodoMVC React example~\cite{todomvc}.
In general, black-box dynamic approaches %
inherently produce fewer false positives and offer precise support for dynamic language features across languages and frameworks.

\para{Grey-box Fuzzing}
Grey-box web fuzzing~\cite{DBLP:conf/esorics/RooijCKPA21, DBLP:journals/corr/abs-2108-08455, DBLP:conf/sp/TrickelPZDVKWBSD23} typically instruments server-side code to provide coverage feedback to the fuzzing engine.
Trickel et al.~\cite{DBLP:conf/sp/TrickelPZDVKWBSD23} apply this to detect SQL and command injection.
This approach is limited to single-shot reflected injection vulnerabilities.
While the fuzzer can induce application changes, this approach does not reason about states.
Gauthier et al.~\cite{DBLP:journals/corr/abs-2108-08455} abstract web applications into REST models, and then use a black-box fuzzer on Node.JS applications.
G{\"u}ler et al.~\cite{DBLP:conf/uss/GulerSSBGXKH24}
add bug oracles to provide feedback in an instrumented PHP interpreter.
In contrast,
our approach aims to instead generate structured inputs by correctly interacting with the client-side interface.

\para{Black-box Scanners}
CrawlJax~\cite{DBLP:conf/icwe/MesbahBD08} infers a state flow graph, including support for client-side UI changes.
This has since been incorporated into the \zap scanner, and is therefore included in our evaluation.
Enemy of the State~\cite{DBLP:conf/uss/DoupeCKV12} introduces another scanner that can model server-side state.

\jak~\cite{DBLP:conf/raid/PellegrinoTBR15} improves detection of possible interactions, such as events, network APIs, and dynamic URLs and forms, through dynamic analysis of JavaScript code.
This analysis through event registration hooking is reused in \bw~\cite{DBLP:conf/sp/ErikssonPS21}, together with
navigation modeling with traversing and tracking inter-state dependencies.
In this paper we evaluate \shortname against \blackostrich, which extends \bw's input validation support.
Our approach improves the crawler's core navigation modeling, crawling strategy, and form handling
increasing both coverage and vulnerability detection.

\evocrawl~\cite{DBLP:conf/ndss/GuoKLL25} utilizes evolutionary search to crawl modern applications.
Their approach improves code coverage and form submissions, as the evolutionary search can improve their navigation strategy.
However, the discovery of interactable elements is limited to a predefined set of tags.
Our method's interactable element discovery could improve their evolutionary crawler's performance.

LOAD-AND-ACT~\cite{DBLP:conf/isw/WeidmannBW23} %
improves client-side coverage to find vulnerabilities in a single page,
as opposed to entire applications.
Their element discovery improves on \jak by removing
non-interactable elements using WebDriver~\cite{webdriver}.
However, this still misses framework elements that use global event listeners on the body.

Reinforcement Learning (RL) based scanners such as the recent work by Cazzaro et al.~\cite{DBLP:conf/dsn/CazzaroCKSP25} are intriguing.
However, we do not include this scanner in our evaluation as its source code is not published.

ReScan~\cite{DBLP:conf/ndss/DrakonakisIP23} is a middleware that aims to enhance any scanner by clustering URLs and prerendering the DOM.
This is helpful for older scanners like w3af that do not support JavaScript.

\para{Creating Database State}
SynthDB~\cite{DBLP:conf/ndss/ChenLC0L23} prepares database-backed PHP web applications for security testing by synthesizing a database after collecting constraints with concolic execution.
Spider-Scents~\cite{DBLP:conf/uss/0001EDS24} find portions of XSS vulnerabilities
by directly inserting payloads into the database.
In contrast, we attempt to create application state, including the database, with the client-side interface.

\para{LLMs for Security Scanning}
\label{sec:yurascanner}
\yurascanner~\cite{DBLP:conf/ndss/StafeevRSKP25}
leverages an LLM for task-driven scanning.
The LLM helps define and execute tasks and workflows, such that deeper states
can be scanned for vulnerabilities.
Their task-driven approach with a goal-based agent is a departure from the more typical crawler approach we assume, and can therefore have complementary results.

Yoon et al.~\cite{DBLP:journals/corr/abs-2311-08649} apply LLM agents to fetch interactable elements for task goals in Android applications.
Future web crawlers could align with the approach of ScreenAI~\cite{DBLP:conf/ijcai/BaechlerSWZMECL24}, where a vision language model identifies UI elements like search bars. %

Hoyen~\cite{DBLP:journals/corr/abs-2504-20801} is another recent work that uses an LLM for both the navigation strategy and form solving in a black-box scanner.
Their use of an LLM to predict user intention aligns with our own goal of exploring intended code paths.
However, we do not evaluate this scanner as its source code has not been published yet.

\para{Input Validation}
\blackostrich~\cite{DBLP:conf/ccs/ErikssonSMRS23} %
focuses on input validation by solving regexes with SMT. %
ExpoSE~\cite{DBLP:journals/corr/abs-1810-05661} is a symbolic execution engine for JS that can solve regex-based constraints.
FormWhisperer~\cite{DBLP:conf/kbse/KartheinSZ24} uses symbolic analysis of HTML and JS that extracts and solves constraints on form inputs, including between form fields.
In contrast, we solve forms with diverse input validation approaches.
However, these approaches are complementary in that regex solving could provide an input for the LLM, or vice versa.

\section{Conclusion}
Our work showcases the challenges modern web applications pose for
state-of-the-art black-box scanners.
To combat these challenges and help developers secure their applications, we
develop a novel method to better interact with these applications.
Our method achieves \technique by improving detection of interactable elements and
ordering interactions with these elements, while using an
LLM-based approach to tackle the diverse input validations in forms.
We implement this method in our scanner \shortname and evaluate it
on \totalApplications web applications. Our results show improvement in both
coverage and vulnerability detection, with an average increase in coverage of
\improvedCoverageVSAllNoDecFROZEN{} compared to the union of all scanners,
 and a total of \totalXSS XSS vulnerabilities across \xssApplications applications.

\bibliographystyle{IEEEtran}
\bibliography{condensed-dblp,custom}

\appendix %

\subsection{Ethical Considerations} %

By actively scanning only our local clones of these applications in a controlled environment, we avoid any harm caused by scanners on the web.
We are handling the discovered security vulnerabilities in accordance with the best practices of ethics in security~\cite{DBLP:journals/ieeesp/BaileyDKM12}.
We have reported our findings to the affected vendors,
following coordinated vulnerability disclosure for all discovered vulnerabilities.
We will include any additional responses from vendors in the final paper.

\subsection{Ablation Study Details}
\label{sec:ablation_study}

Here, we present details of the two ablation studies to determine the effects of the LLM-powered form solving  and element discovery methods used in \shortname, previously summarized in \Cref{sec:ablation_study_summary}.
The LLM-ablated version \ablatedLLM instead uses the standard form input method from the underlying \bw scanner.
\bw, and therefore \ablatedLLM, apply heuristics based on input types, similar to \zap.
The element-ablated version \ablatedELM instead reverts back to the element discovery of \bw.
The \bw scanner relies on semantic elements and JavaScript event listeners, %
similar to \yura and \jak.
We evaluate both ablations on the same applications and metrics.

\para{Coverage}
In \Cref{table:ablation} we present the coverage achieved by both \shortname and
the ablated versions \ablatedLLM and \ablatedELM.
On average, adding the LLM or element discovery method improves coverage by \improvedCoverageVSbwinputs and \improvedCoverageVSbwelements respectively.
While \shortname outperforms \ablatedLLM and \ablatedELM, there %
are notable differences in their relative performance.

We start by contrasting \shortname and \ablatedLLM, presented in \Cref{table:ablation}, to show when advanced form solving is necessary.
For applications that require solving complex forms early, such as \kanboard, \hotcrp, and \leantime,
the LLM method shows a clear benefit.
Coverage is otherwise limited, as a scanner cannot create a project and task in \kanboard or \leantime, or submit a paper in \hotcrp.
We also note that using an LLM to find the correct submission buttons in forms (see \Cref{llm-crawl-subprompts}) helps in cases such as \tinyfilemanager where the first button is atypically used for ``cancel''.
For applications like \oscommerce there is a smaller difference in coverage, which we discuss in \Cref{sec:oscommerce_coverage}.

The \ablatedELM version also generally performs worse than \shortname
because the scanner is not able to
detect elements, and is mistakenly detecting non-interactable elements.
For example, \piwigo use \texttt{label} tags for buttons and
define the event listener on the body, using jQuery, instead of on the element.
Therefore, the \ablatedELM does not find it, leading to a decrease in coverage, even compared to \ablatedLLM.
Furthermore, as the previous method for detecting elements does not consider whether they are hidden or occluded,
the ablated version often tries to interact with elements in the wrong state, resulting in decreased coverage.
For example, clicking a submenu item before opening the menu.

\para{Vulnerabilities}
The ablated \ablatedLLM finds \ablatedLLMTotalXSS XSS across \ablatedLLMXSSApplications applications.
We note a large overlap of \ablatedLLMOverlapXSS vulnerabilities between \shortname and the ablated \ablatedLLM.
Similarly for \ablatedELM, this ablation finds \ablatedELMTotalXSS XSS across \ablatedELMXSSApplications applications with
an overlap of \ablatedELMOverlapXSS.
While both ablations find many vulnerabilities in \oscommerce, their XSS count is further inflated
by a single form in \leantime with 7 XSS vulnerabilities.
There is no theoretical reason that prevents \shortname from finding these.
In fact, \shortname does find this form and its XSS in the repeated evaluation case study in \Cref{sec:repeated_evaluation}.
This indicates that more time or repeated runs are helpful to minimize false negatives, as discussed
in \Cref{sec:false_negatives}.
Most notably, neither ablated version finds XSS in as many applications as \shortname.

Some clear examples of why ablated versions fail are on \piwigo and \kanboard.
\piwigo's unique button type, using \texttt{label} elements, makes it impossible
for \ablatedELM to click the button and generate the necessary form for XSS.
Conversely, complex forms on \kanboard make it impossible for \ablatedLLM
to generate the correct inputs to create new projects and tasks.

\begin{table}[h!]
	\caption{LoC executed on the server by \shortname and the ablated versions \ablatedLLM and \ablatedELM.}
	\label{table:ablation}
	\begin{center}
		\begin{adjustbox}{max width=\textwidth}
			\begin{tabular}{  l |  r r r     }
				\toprule
				& \shortname & \ablatedLLM & \ablatedELM
				\\ \midrule

				\dokuwiki                      &  \textbf{\numprint{18686}}      &  \numprint{18022}               &  \numprint{16483}               \\
				\hotcrp                        &  \textbf{\numprint{36429}}      &  \numprint{21255}               &  \numprint{25782}               \\
				\kanboard                      &  \textbf{\numprint{16668}}      &  \numprint{7286}                &  \numprint{13638}               \\
				\leantime                      &  \textbf{\numprint{29262}}      &  \numprint{25122}               &  \numprint{28608}               \\
				\nextcloud                     &  \textbf{\numprint{67599}}      &  \numprint{65617}               &  \numprint{62352}               \\
				\oscommerce                    &  \textbf{\numprint{121048}}     &  \numprint{101423}              &  \numprint{77208}               \\
				\piwigo                        &  \textbf{\numprint{50996}}      &  \numprint{23739}               &  \numprint{22454}               \\
				\tinyfilemanager               &  \textbf{\numprint{1423}}       &  \numprint{1170}                &  \numprint{1197}                \\
				\wordpress                     &  \textbf{\numprint{59732}}      &  \numprint{57740}               &  \numprint{56796}               \\

				\bottomrule
			\end{tabular}
		\end{adjustbox}
	\end{center}
\end{table}

\subsection{Crawlers v.s. Task Solvers}
\label{appendix:task_solvers}
An important difference between a crawler and a task solver is their respective recurrent loop or single-pass run.
The recurrent crawling loop of \shortname allows it to exercise deeper application functionality that is only uncovered after some actions, rather than being immediately available.
On the other hand, task solvers such as \yurascanner first make a static list of tasks, and then execute those tasks.
\yurascanner uses a limited crawl as the context for an LLM to generate a list of tasks.
While an LLM could possibly infer that some deeper functionality will appear, it often fails to in practice.
For example, on \hotcrp, the tasks \yurascanner generates correspond primarily to surface-level actions.
Deeper functionalities in the application, such as writing a review, are not included in these tasks, as this functionality only appears after a successful paper submission.
As can be seen in \Cref{sec:hotcrp_states}, while \yurascanner is successful at performing these shallow actions, it misses deeper states of \hotcrp.

\subsection{LLM Form Solving Implementation and Prompt}
\label{sec:llm_form_details}

We provide additional details about the implementation and prompt for our LLM form solving method here.

\para{Function Calling}
To describe the task of form solving and easily parse model output, we provide an API for the LLM.
This function is
called by the LLM to input the result of a form solve,
instructing the LLM to provide the structured output as a list of (CSS locator, user input) pairs.

\para{Input Automation}
The (CSS locator, user input) pairs are read from the LLM's output, and input is automatically entered into the form elements.
This input automation includes support for clicking on checkbox and radio types, choosing appropriate values from select elements, uploading files, and typing into text fields.

\para{Few-shot Examples}
The LLM is also provided additional context about the problem domain in the form of six simple examples of forms and function calls the LLM should generate to solve them.

\para{Prompt}
\Cref{fig:llm_template} contains the complete prompting template for LLM form solving.
Note that the instruction ``Do make assumptions about what values to plug
into functions'' is deliberate since the model should assume plausible
values and proceed.
This prompt is used as described in \Cref{sec:llm-implementation}.
\begin{figure}
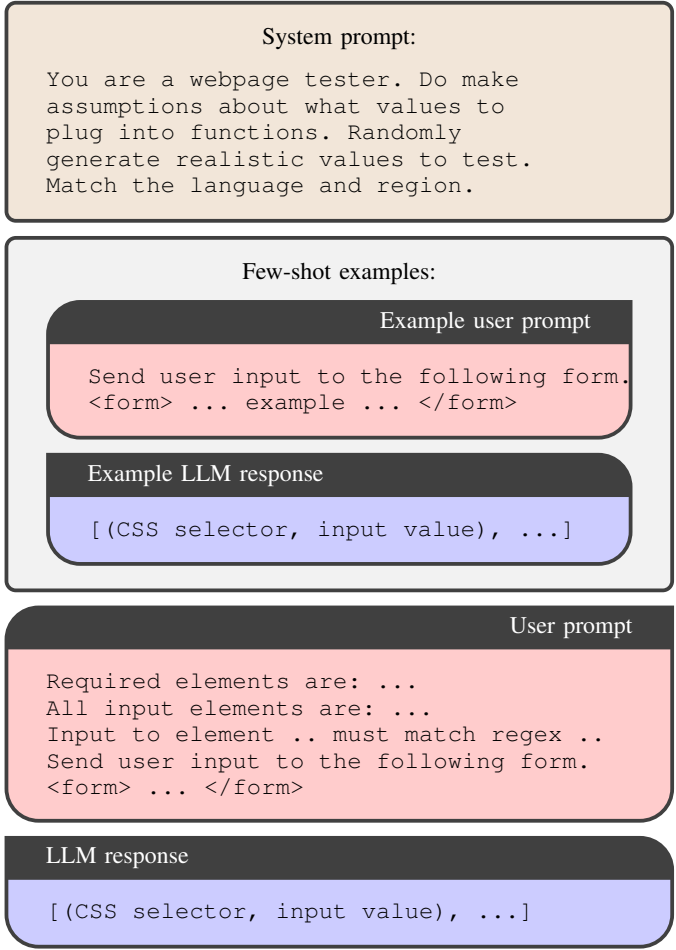

\AtBeginEnvironment{tcolorbox}{\small}
\begin{tcolorbox}[colback=cl2]
\centering
System prompt:
\begin{verbatim}
You are a webpage tester. Do make
assumptions about what values to
plug into functions. Randomly
generate realistic values to test.
Match the language and region.
\end{verbatim}
\end{tcolorbox}

\begin{tcolorbox}
\centering
Few-shot examples:
\begin{tcolorbox}[sharp corners=northeast,arc=4mm,colback=cl1,title={Example user prompt},halign title=flush right]
\begin{verbatim}
Send user input to the following form.
<form> ... example ... </form>
\end{verbatim}
\end{tcolorbox}
\begin{tcolorbox}[sharp corners=northwest,arc=4mm,colback=cl3,title=Example LLM response]
\begin{verbatim}
[(CSS selector, input value), ...]
\end{verbatim}
\end{tcolorbox}
\end{tcolorbox}

\begin{tcolorbox}[sharp corners=northeast,arc=4mm,colback=cl1,title={User\ prompt},halign title=flush right]
\begin{verbatim}
Required elements are: ...
All input elements are: ...
Input to element .. must match regex ..
Send user input to the following form.
<form> ... </form>
\end{verbatim}
\end{tcolorbox}
\begin{tcolorbox}[sharp corners=northwest,arc=4mm,colback=cl3,title=LLM response]
\begin{verbatim}
[(CSS selector, input value), ...]
\end{verbatim}
\end{tcolorbox}

\caption{Prompting template for LLM form solving}
\label{fig:llm_template}
\end{figure}

\subsection{LLM Parameters and Cost}
\label{appendix:llm_details}

In our evaluation, we use the Gemini 2.5 Flash LLM with a temperature of 1.7.
We did not systematically evaluate other choices for these parameters.
While this can be improved in future work, integrating more careful selection of which form inputs to provide, as indicated by \Cref{sec:tinyfilemanager_states} and \Cref{sec:false_negatives} may be more beneficial.

We do offer a small 8-hour evaluation of the performance of the scanner across some different Gemini models and temperatures, on \tinyfilemanager (\Cref{tab:llm_tinyfilemanager}) and \kanboard (\Cref{tab:llm_kanboard}).
The chosen model and temperature perform well on \tinyfilemanager, with only marginal gains from switching to the 2.5 Flash-Lite model.
However, on \kanboard, there is much more variance in how different models and temperatures perform, which can indicate the potential improvement from a systematic parameter search, including model and temperature.
We note that the majority of (model, temperature) parametrizations explore different parts of the application, with 19291 different LoC discovered by all parametrizations in this table.
Only a few particularly successful parametrizations, such as (2.5 Pro, 0.0), (2.5 Flash-Lite, 0.0), and (2.5 Flash-Lite, 2.0) discover most of this surface.

\begin{table}
	\caption{
		Scanner performance on \tinyfilemanager with different models and temperatures, compared to the evaluated 2.5 Flash model with temperature 1.7.
	}
	\label{tab:llm_tinyfilemanager}
	\begin{center}
		\begin{adjustbox}{max width=0.48\textwidth}
		\begin{tabular}{l | rrr | rrr | rrr }
		\toprule
		Model     & \multicolumn{3}{c|}{2.5 Flash-Lite}  & \multicolumn{3}{c|}{2.5 Flash}  & \multicolumn{3}{c}{2.5 Pro} \\
		Temperature & $A \setminus B$ & $A \cap B$   & $B \setminus A$
				& $A \setminus B$ & $A \cap B$   & $B \setminus A$
				& $A \setminus B$ & $A \cap B$   & $B \setminus A$ \\
		\midrule
0.0 & 98 & 1317 & 4 & 103 & 1312 & 15 & 102 & 1313 & 6 \\
1.0 & 9 & 1406 & 11 & 144 & 1271 & 12 & 100 & 1315 & 16 \\
1.7 & 0 & 1415 & 6 & 0 & 1415 & 0 & 27 & 1388 & 9 \\
2.0 & 97 & 1318 & 9 & 98 & 1317 & 6 & 97 & 1318 & 5 \\
		\bottomrule
		\end{tabular}
		\end{adjustbox}
	\end{center}
\end{table}

\begin{table}
	\caption{
		Scanner performance on \kanboard with different models and temperatures, compared to the evaluated 2.5 Flash model with temperature 1.7.
	}
	\label{tab:llm_kanboard}
	\begin{center}
		\begin{adjustbox}{max width=0.48\textwidth}
		\begin{tabular}{l | rrr | rrr | rrr }
		\toprule
		Model     & \multicolumn{3}{c|}{2.5 Flash-Lite}  & \multicolumn{3}{c|}{2.5 Flash}  & \multicolumn{3}{c}{2.5 Pro} \\
		Temperature & $A \setminus B$ & $A \cap B$   & $B \setminus A$
				& $A \setminus B$ & $A \cap B$   & $B \setminus A$
				& $A \setminus B$ & $A \cap B$   & $B \setminus A$ \\
		\midrule
0.0 & 690 & 7682 & 9593 & 149 & 8223 & 1651 & 1556 & 6816 & 10112 \\
1.0 & 1208 & 7164 & 1340 & 1540 & 6832 & 1602 & 1186 & 7186 & 3254 \\
1.7 & 343 & 8029 & 1421 & 0 & 8372 & 0 & 89 & 8283 & 2447 \\
2.0 & 703 & 7669 & 9618 & 665 & 7707 & 1712 & 1001 & 7371 & 5584 \\
		\bottomrule
		\end{tabular}
		\end{adjustbox}
	\end{center}
\end{table}

We present the LLM usage details of \shortname, in terms of token usage and cost, in \Cref{tab:llm_cost}.

\begin{table*}
	\caption{
		Token usage and cost of LLM usage during an 8-hour evaluation run of \shortname with the Gemini 2.5 Flash LLM.
		Costs are calculated using Google Cloud's pricing model~\cite{geminipricing}.
		A cost based solely on number of I/O tokens is first presented, and then the lowered cost when repeated inputs (i.e., the prompt template including examples) are cached and billed at a cheaper rate.
	}
	\label{tab:llm_cost}
	\begin{center}
		\begin{tabular} {l | r | r | r | r | r | r | r}
			\toprule
			Application & API calls & Input & Cached input & Output & Total & Cost w/o cache & Cost w/ cache \\
			\midrule
			\dokuwiki & 157 & 648888 & 78651 & 6127 & 706518 & \$0.87 & \$0.78 \\
			\hotcrp & 1022 & 4429578 & 749968 & 57404 & 4898428 & \$6.11 & \$5.36 \\
			\kanboard & 358 & 858286 & 254407 & 15621 & 991104 & \$1.23 & \$0.97 \\
			\leantime & 922 & 6354132 & 730564 & 35743 & 6740550 & \$8.30 & \$7.57 \\
      		\nextcloud & 77 & 380593 & 38922 & 2471 & 410489 & \$0.50 & \$0.46 \\
			\oscommerce & 324 & 2263477 & 202604 & 22115 & 2490902 & \$3.05 & \$2.85 \\
			\piwigo & 49 & 197600 & 11042 & 1750 & 226661 & \$0.26 & \$0.25 \\
			\tinyfilemanager & 45 & 99701 & 25703 & 1570 & 116146 & \$0.14 & \$0.11 \\
			\wordpress & 140 & 643265 & 101309 & 4849 & 701640 & \$0.85 & \$0.74 \\
			\midrule
			Total & 3094 & 15875520 & 2193170 & 147650 & 17282438 & \$21.32 & \$19.10 \\
			\bottomrule
		\end{tabular}
	\end{center}
\end{table*}

\subsection{Web Applications}
\label{appendix:web_apps}
In \Cref{tab:web_applications} we present the exact version of each web application used in the evaluation, as well as the amount of stars their code repositories have on GitHub, and which prior work has also evaluated on any version of these applications.
In \Cref{tab:frontend-frameworks} we present the front-end JavaScript frameworks used by the evaluated web applications.

\begin{table*}
	\caption{
		The version, number of GitHub stars, and prior work for each web application. %
	}
	\label{tab:web_applications}
	\begin{center}
		\begin{tabular}{  l | l | l | l | l | l}
			\toprule
			Application       & Version & Date & Description & GitHub stars & Prior work \\
			\midrule
			\dokuwiki         & 2024-02-06b & 2024-02-06 & CMS & 4.6k & \cite{DBLP:conf/ndss/GuoKLL25, DBLP:conf/uss/GulerSSBGXKH24} \\
			\hotcrp           & 9588ab0 & 2024-09-18 & Conference Manager & 410 & \cite{DBLP:conf/sp/ErikssonPS21, DBLP:conf/uss/DahseH14, DBLP:conf/ndss/GuoKLL25, DBLP:conf/uss/GulerSSBGXKH24} \\
			\kanboard         & 1.2.40 & 2024-09-26 & Project Manager & 9.6k & \cite{DBLP:conf/ndss/GuoKLL25, DBLP:conf/uss/GulerSSBGXKH24} \\
			\leantime         & 3.3.3 & 2024-11-27 & Project Manager & 9.5k & \cite{DBLP:conf/ndss/StafeevRSKP25} \\
			\nextcloud & 33.0.0.16 & 2026-02-17 & Collaboration Platform & 43.7k & \cite{DBLP:conf/uss/GulerSSBGXKH24} \\
			\oscommerce       & 4.14.63493 & 2024-04-26 & Ecommerce & 54 & \cite{DBLP:conf/sp/ErikssonPS21, DBLP:conf/ndss/ChenLC0L23, DBLP:conf/uss/DahseH14, DBLP:journals/corr/abs-2504-20801} \\
			\piwigo           & 14.3.0 & 2024-03-01 & Photo Album & 3.8k & \cite{DBLP:conf/uss/0001EDS24, DBLP:conf/raid/PellegrinoTBR15, DBLP:conf/uss/GulerSSBGXKH24} \\
			\tinyfilemanager  & 2.6 & 2024-11-05 & File Explorer & 5.9k & N/A \\
			\wordpress        & 6.6.2 & 2024-09-10 & CMS & 21.1k & \cite{DBLP:conf/sp/ErikssonPS21, DBLP:conf/ndss/ChenLC0L23, DBLP:conf/uss/DoupeCKV12, DBLP:conf/uss/0001EDS24, DBLP:conf/raid/PellegrinoTBR15, DBLP:conf/uss/StafeevP24, DBLP:conf/esorics/RooijCKPA21, DBLP:conf/ndss/GuoKLL25, DBLP:conf/uss/GulerSSBGXKH24, DBLP:journals/corr/abs-2504-20801} \\
			\bottomrule
		\end{tabular}
	\end{center}
\end{table*}

\begin{table}
\caption{ Front-end JS frameworks used by the PHP web applications included in the evaluation. }
\label{tab:frontend-frameworks}
\centering
\begin{tabular}{ l | l }
\toprule
Application & Front-end frameworks \\ %
\midrule
\dokuwiki & jQuery \\ %
\hotcrp & jQuery \\ %
\kanboard & jQuery \\ %
\leantime & Bootstrap, jQuery \\ %
\nextcloud & jQuery, Vue.js \\ %
\oscommerce & Bootstrap, jQuery \\ %
\piwigo & jQuery \\ %
\tinyfilemanager & Bootstrap, jQuery \\ %
\wordpress & jQuery, React \\ %
\bottomrule
\end{tabular}
\end{table}

We modify the applications to help scanners automatically authenticate, either by instrumenting the authentication function or by simulating an automatic login.
This is necessary as many of the other scanners we evaluate fail to correctly sign in on some of these modern applications.
It also helps the scanners re-authenticate even if they sign out by mistake. Avoiding sign-outs is an orthogonal problem (see ReScan~\cite{DBLP:conf/ndss/DrakonakisIP23}), but one that all scanners face.
With these modifications we can learn more about the crawling capabilities of the various scanners without getting stuck on their varying support for successful authentication.

Additionally, since we are interested in detecting vulnerable \textit{code} in web applications, we deactivate hardening functionality such as Content-Security Policy (CSP). In practice, this only affects \kanboard.
We argue that it is still important to find XSS vulnerabilities even if they are mitigated by CSP as someone could run an instance without CSP.
Furthermore, there is a precedent for the \kanboard developers fixing XSS vulnerabilities~\cite{DBLP:conf/ndss/GuoKLL25}.

\subsection{Coverage Metrics}
\label{appendix:discussion_metric}
\label{sec:discussion_metric}

Server-side code coverage, such as that gathered through xDebug in PHP web applications, is the most common way to measure coverage in a web application scanning or fuzzing session.
However, other types of coverage can be measured.
For example, the recent SoK on web security crawlers~\cite{DBLP:conf/uss/StafeevP24} measures not only (1) server-side code coverage, but also (2) JavaScript source coverage, and (3) link coverage.
JavaScript source coverage is defined by collecting the hashes of retrieved inline or external JavaScript.

JavaScript code coverage (4) can also be measured by retrieving the statistics of total and unused bytes of JavaScript code through the Chrome DevTools.
This has been used by Kang et al.~\cite{DBLP:conf/ndss/Kang0C22}, but has not yet been widely adopted by other scanning works.
Comparative evaluation is difficult, as each scanner needs to gather such metrics.

Therefore, we only present the coverage in the (1) metric: server-side code coverage.
Possible client-side metrics for coverage (2-4) are harder to collect.
Dynamically generated scripts and links will perturb these measurements.

While a notion of \emph{state coverage} would best measure the ability for scanners to reach deep states in web applications, this metric is nontrivial to define.
Other scanners that target deep state, such as \yurascanner, still only measure coverage with standard metrics.
In a database-backed application without any other state, \emph{database coverage} might approximate this~\cite{DBLP:conf/uss/0001EDS24}.
Even defining states is tricky.
\yurascanner takes a first step by manually categorizing states from steps in task traces, but can only use this to identify task success or failure.
Similarly, the depth of traces, either in tasks or crawling, is hard to measure. %
Measuring this requires knowing the shortest possible path to any given application state.
The approximation \yurascanner takes by counting the number of steps in a task can be misleading, as steps can, and do, repeat.
We present our own analysis of state depth in \Cref{sec:state_depth}.

\subsection{Client-side Coverage Data}

In \Cref{table:clientside_compare_blackostrich} we present the byte-level client-side coverage achieved by \shortname and \blackostrich, as described in \Cref{sec:clientside_coverage}.

\begin{table}
	\caption{
		Comparison of client-side coverage in executed KBs of Javascript (with percent of total in parentheses) between \shortname and \blackostrich. The \textit{Total} column indicates the \texttt{totalBytes} as reported by Chrome.	}
	\label{table:clientside_compare_blackostrich}
	\begin{center}
		\begin{adjustbox}{max width=0.48\textwidth}
			\begin{tabular}{  l |  r r r r }
				\toprule
				Crawlers        & \multicolumn{4}{c}{\shortname ($A$) v.s. \blackostrich ($B$)}  \\
				               & $A \setminus B$ & $A \cap B$   & $B \setminus A$	& Total (KB) \\
			    \midrule

\dokuwiki                      & \textbf{\numprint{360}} (45\%) & \numprint{152} (19\%)          & \numprint{13} (2\%)            & \numprint{805}                \\
\hotcrp                        & \textbf{\numprint{517}} (38\%) & \numprint{210} (15\%)          & \numprint{53} (4\%)            & \numprint{1371}               \\
\kanboard                      & \textbf{\numprint{23}} (2\%)   & \numprint{276} (27\%)          & \numprint{7} (1\%)             & \numprint{1027}               \\
\leantime                      & \textbf{\numprint{468}} (7\%)  & \numprint{2046} (32\%)         & \numprint{213} (3\%)           & \numprint{6472}               \\
\nextcloud                     & \textbf{\numprint{33663}} (35\%) & \numprint{26211} (27\%)        & \numprint{1133} (1\%)          & \numprint{96373}              \\
\oscommerce                    & \textbf{\numprint{38361}} (74\%) & \numprint{1242} (2\%)          & \numprint{3233} (6\%)          & \numprint{51951}              \\
\piwigo                        & \textbf{\numprint{1898}} (38\%) & \numprint{179} (4\%)           & \numprint{6} (0\%)             & \numprint{4965}               \\
\tinyfilemanager               & \textbf{\numprint{283}} (28\%) & \numprint{268} (27\%)          & \numprint{1} (0\%)             & \numprint{1002}               \\
\wordpress                     & \textbf{\numprint{6327}} (43\%) & \numprint{3835} (26\%)         & \numprint{254} (2\%)           & \numprint{14670}              \\

				\bottomrule
				\end{tabular}
			\end{adjustbox}
		\end{center}
	\end{table}

\subsection{Non-semantic Interactable Elements Data}

In \Cref{table:nonsemantic_interactable} we present the elements, non-semantic elements, and non-semantic tags found by \shortname, as described in \Cref{sec:nonsemantic_interactable}.

\begin{table}
	\centering
	\caption{Interactable elements, non-semantic elements, and non-semantic tags found by \shortname. }
	\label{table:nonsemantic_interactable}
	\begin{adjustbox}{max width=0.48\textwidth}
		\begin{tabular}{l | rrrr  }

			\toprule
			Scanner     & \multicolumn{3}{c}{\shortname}    \\
			Application & Elements & Elements (NS) & Tags (NS)     \\
			\midrule

			\dokuwiki             & 2000       & 26         & 7          \\
			\hotcrp               & 4776       & 21         & 5          \\
			\kanboard             & 2662       & 96         & 7          \\
			\leantime             & 3402       & 158        & 10         \\
			\nextcloud            & 7853       & 2106       & 17         \\
			\oscommerce           & 4629       & 3502       & 13         \\
			\piwigo               & 2234       & 174        & 9          \\
			\tinyfilemanager      & 797        & 6          & 2          \\
			\wordpress            & 2789       & 258        & 8          \\

			\bottomrule
		\end{tabular}
	\end{adjustbox}
\end{table}

\subsection{Coverage Results Details}
\label{appendix:coverage_results}
In \Cref{table:locs,table:locs_sota} we present the exact numbers for the coverage evaluation.
In \Cref{table:coverage_improvement} we present the coverage improvement over other scanners or their union, both on individual applications and averaged.

\begin{table*}%
	\caption{ Coverage improvement by \shortname{} over another crawler ($(|A|-|B|)/(|B|)$). }
	\label{table:coverage_improvement}
	\begin{center}
		\begin{adjustbox}{max width=\textwidth}
			\begin{tabular}{ l |  r |  r  | r  |  r | r | r | r}
				\toprule
				Crawler & \arachni & \blackostrich & \wapiti & \zap & \yura & \evocrawl & Union \\
        \midrule
        \dokuwiki & \improvedCoverageTwoONdokuwikiVSarachni & \improvedCoverageTwoONdokuwikiVSblackostrich & \improvedCoverageTwoONdokuwikiVSwapiti & \improvedCoverageTwoONdokuwikiVSzap & \improvedCoverageTwoONdokuwikiVSyura & \improvedCoverageTwoONdokuwikiVSevocrawl & \improvedCoverageVsAllOndokuwiki\\
        \hotcrp & \improvedCoverageTwoONhotcrpVSarachni & \improvedCoverageTwoONhotcrpVSblackostrich & \improvedCoverageTwoONhotcrpVSwapiti & \improvedCoverageTwoONhotcrpVSzap & \improvedCoverageTwoONhotcrpVSyura & \improvedCoverageTwoONhotcrpVSevocrawl & \improvedCoverageVsAllOnhotcrp\\
        \kanboard & \improvedCoverageTwoONkanboardVSarachni & \improvedCoverageTwoONkanboardVSblackostrich & \improvedCoverageTwoONkanboardVSwapiti & \improvedCoverageTwoONkanboardVSzap & \improvedCoverageTwoONkanboardVSyura & \improvedCoverageTwoONkanboardVSevocrawl & \improvedCoverageVsAllOnkanboard\\
        \leantime & \improvedCoverageTwoONleantimeVSarachni & \improvedCoverageTwoONleantimeVSblackostrich & \improvedCoverageTwoONleantimeVSwapiti & \improvedCoverageTwoONleantimeVSzap & \improvedCoverageTwoONleantimeVSyura & \improvedCoverageTwoONleantimeVSevocrawl & \improvedCoverageVsAllOnleantime\\
        \nextcloud & \improvedCoverageTwoONnextcloudVSarachni & \improvedCoverageTwoONnextcloudVSblackostrich & \improvedCoverageTwoONnextcloudVSwapiti & \improvedCoverageTwoONnextcloudVSzap & \improvedCoverageTwoONnextcloudVSyura & \improvedCoverageTwoONnextcloudVSevocrawl & \improvedCoverageVsAllOnnextcloud\\
        \oscommerce & \improvedCoverageTwoONoscommerceVSarachni & \improvedCoverageTwoONoscommerceVSblackostrich & \improvedCoverageTwoONoscommerceVSwapiti & \improvedCoverageTwoONoscommerceVSzap & \improvedCoverageTwoONoscommerceVSyura & \improvedCoverageTwoONoscommerceVSevocrawl & \improvedCoverageVsAllOnoscommerce\\
        \piwigo & \improvedCoverageTwoONpiwigoVSarachni & \improvedCoverageTwoONpiwigoVSblackostrich & \improvedCoverageTwoONpiwigoVSwapiti & \improvedCoverageTwoONpiwigoVSzap & \improvedCoverageTwoONpiwigoVSyura & \improvedCoverageTwoONpiwigoVSevocrawl & \improvedCoverageVsAllOnpiwigo\\
        \tinyfilemanager & \improvedCoverageTwoONtinyfilemanagerVSarachni & \improvedCoverageTwoONtinyfilemanagerVSblackostrich & \improvedCoverageTwoONtinyfilemanagerVSwapiti & \improvedCoverageTwoONtinyfilemanagerVSzap & \improvedCoverageTwoONtinyfilemanagerVSyura & \improvedCoverageTwoONtinyfilemanagerVSevocrawl & \improvedCoverageVsAllOntinyfilemanager\\
        \wordpress & \improvedCoverageTwoONwordpressVSarachni & \improvedCoverageTwoONwordpressVSblackostrich & \improvedCoverageTwoONwordpressVSwapiti & \improvedCoverageTwoONwordpressVSzap & \improvedCoverageTwoONwordpressVSyura & \improvedCoverageTwoONwordpressVSevocrawl & \improvedCoverageVsAllOnwordpress\\
				\midrule
				Mean & \improvedCoverageVSarachni & \improvedCoverageVSblackostrich & \improvedCoverageVSwapiti & \improvedCoverageVSzap & \improvedCoverageVSyura & \improvedCoverageVSevocrawl & \improvedCoverageVSAll \\
				\bottomrule
			\end{tabular}
		\end{adjustbox}
	\end{center}
\end{table*}

\begin{table*}%
	\caption{
		LoC executed on the server.
		Each column represents the comparison between \shortname{} and another crawler.
		The cells contain three numbers:
		unique LoC covered by \shortname{} ($A \setminus B$),
		LoC covered by both crawlers ($A \cap B$)
		and unique LoC covered by the other crawler ($B \setminus A$).
		The numbers in bold highlight which crawler has the best coverage.	}
	\label{table:locs}
	\begin{center}
		\begin{adjustbox}{max width=\textwidth}
			\begin{tabular}{  l |  r r r  |  r r r  |  r r r  |  r r r }
				\toprule
				Crawler        & \multicolumn{3}{c|}{Arachni}      & \multicolumn{3}{c|}{Black Ostrich }      & \multicolumn{3}{c|}{Wapiti }            & \multicolumn{3}{c}{ZAP}            \\
				& $A \setminus B$ & $A \cap B$   & $B \setminus A$
				& $A \setminus B$ & $A \cap B$   & $B \setminus A$
				& $A \setminus B$ & $A \cap B$   & $B \setminus A$
				& $A \setminus B$ & $A \cap B$   & $B \setminus A$
				\\ \midrule

				\dokuwiki

& \textbf{\numprint{11388}}  & \numprint{7298}     & \numprint{57} %
& \textbf{\numprint{3123}}  & \numprint{15563}     & \numprint{49} %
& \textbf{\numprint{3691}}  & \numprint{14995}     & \numprint{1589} %
& \textbf{\numprint{2190}}  & \numprint{16496}     & \numprint{1495} %
\\
\hotcrp

& \textbf{\numprint{32169}}  & \numprint{4260}     & \numprint{47} %
& \textbf{\numprint{17024}}  & \numprint{19405}     & \numprint{414} %
& \textbf{\numprint{32144}}  & \numprint{4285}     & \numprint{16} %
& \textbf{\numprint{29517}}  & \numprint{6912}     & \numprint{278} %
\\
\kanboard

& \textbf{\numprint{10253}}  & \numprint{6415}     & \numprint{161} %
& \textbf{\numprint{7932}}  & \numprint{8736}     & \numprint{272} %
& \textbf{\numprint{12460}}  & \numprint{4208}     & \numprint{14} %
& \textbf{\numprint{7827}}  & \numprint{8841}     & \numprint{245} %
\\
\leantime

& \textbf{\numprint{9032}}  & \numprint{20230}     & \numprint{1106} %
& \textbf{\numprint{3918}}  & \numprint{25344}     & \numprint{1306} %
& \textbf{\numprint{7510}}  & \numprint{21752}     & \numprint{1078} %
& \textbf{\numprint{3816}}  & \numprint{25446}     & \numprint{1495} %
\\
\nextcloud

& \textbf{\numprint{53412}}  & \numprint{14187}     & \numprint{110} %
& \textbf{\numprint{8489}}  & \numprint{59110}     & \numprint{1360} %
& \textbf{\numprint{54657}}  & \numprint{12942}     & \numprint{16} %
& \textbf{\numprint{22586}}  & \numprint{45013}     & \numprint{2595} %
\\
\oscommerce

& \textbf{\numprint{94109}}  & \numprint{26939}     & \numprint{1701} %
& \textbf{\numprint{42883}}  & \numprint{78165}     & \numprint{3536} %
& \textbf{\numprint{113203}}  & \numprint{7845}     & \numprint{3} %
& \textbf{\numprint{110991}}  & \numprint{10057}     & \numprint{3} %
\\
\piwigo

& \textbf{\numprint{27843}}  & \numprint{23153}     & \numprint{1445} %
& \textbf{\numprint{28543}}  & \numprint{22453}     & \numprint{1021} %
& \textbf{\numprint{46200}}  & \numprint{4796}     & \numprint{4} %
& \textbf{\numprint{39157}}  & \numprint{11839}     & \numprint{655} %
\\
\tinyfilemanager

& \textbf{\numprint{517}}  & \numprint{906}     & \numprint{63} %
& \textbf{\numprint{445}}  & \numprint{978}     & \numprint{34} %
& \textbf{\numprint{513}}  & \numprint{910}     & \numprint{41} %
& \numprint{137}  & \numprint{1286}     & \textbf{\numprint{151}} %
\\
\wordpress

& \textbf{\numprint{17529}}  & \numprint{42203}     & \numprint{2390} %
& \textbf{\numprint{8441}}  & \numprint{51291}     & \numprint{1308} %
& \textbf{\numprint{40228}}  & \numprint{19504}     & \numprint{62} %
& \textbf{\numprint{20325}}  & \numprint{39407}     & \numprint{1013} %
\\

				\bottomrule
			\end{tabular}
		\end{adjustbox}
	\end{center}
\end{table*}

\begin{table*}%
	\caption{
		LoCs executed on the server, compared between \shortname and the latest SOTA crawlers following the format of \Cref{table:locs}. }
	\label{table:locs_sota}
	\begin{center}
		\begin{adjustbox}{max width=\textwidth}
			\begin{tabular}{  l |  r r r  |  r r r     }
				\toprule
				Crawler        & \multicolumn{3}{c|}{\evocrawl}      & \multicolumn{3}{c}{\yurascanner }      \\
				& $A \setminus B$ & $A \cap B$   & $B \setminus A$
				& $A \setminus B$ & $A \cap B$   & $B \setminus A$
				\\ \midrule

				\dokuwiki

& \textbf{\numprint{3995}}  & \numprint{14691}     & \numprint{1642} %
& \textbf{\numprint{3330}}  & \numprint{15356}     & \numprint{692} %
\\
\hotcrp

& \textbf{\numprint{15823}}  & \numprint{20606}     & \numprint{562} %
& \textbf{\numprint{23474}}  & \numprint{12955}     & \numprint{70} %
\\
\kanboard

& \textbf{\numprint{9773}}  & \numprint{6895}     & \numprint{102} %
& \textbf{\numprint{10439}}  & \numprint{6229}     & \numprint{63} %
\\
\leantime

& \textbf{\numprint{3715}}  & \numprint{25547}     & \numprint{1146} %
& \textbf{\numprint{10644}}  & \numprint{18618}     & \numprint{637} %
\\
\nextcloud

& \textbf{\numprint{10153}}  & \numprint{57446}     & \numprint{3656} %
& \textbf{\numprint{15971}}  & \numprint{51628}     & \numprint{6089} %
\\
\oscommerce

& \textbf{\numprint{37104}}  & \numprint{83944}     & \numprint{2980} %
& \textbf{\numprint{42158}}  & \numprint{78890}     & \numprint{5378} %
\\
\piwigo

& \textbf{\numprint{36461}}  & \numprint{14535}     & \numprint{638} %
& \textbf{\numprint{34125}}  & \numprint{16871}     & \numprint{204} %
\\
\tinyfilemanager

& \textbf{\numprint{133}}  & \numprint{1290}     & \numprint{35} %
& \textbf{\numprint{515}}  & \numprint{908}     & \numprint{50} %
\\
\wordpress

& \textbf{\numprint{12584}}  & \numprint{47148}     & \numprint{1443} %
& \textbf{\numprint{16142}}  & \numprint{43590}     & \numprint{64} %
\\

				\bottomrule
			\end{tabular}
		\end{adjustbox}
	\end{center}
\end{table*}

\subsection{\tinyfilemanager State Annotations}
\label{appendix:tinyfilemanager}

\begin{figure}
	\centering
	\includegraphics[angle=90,width=0.4\textwidth]{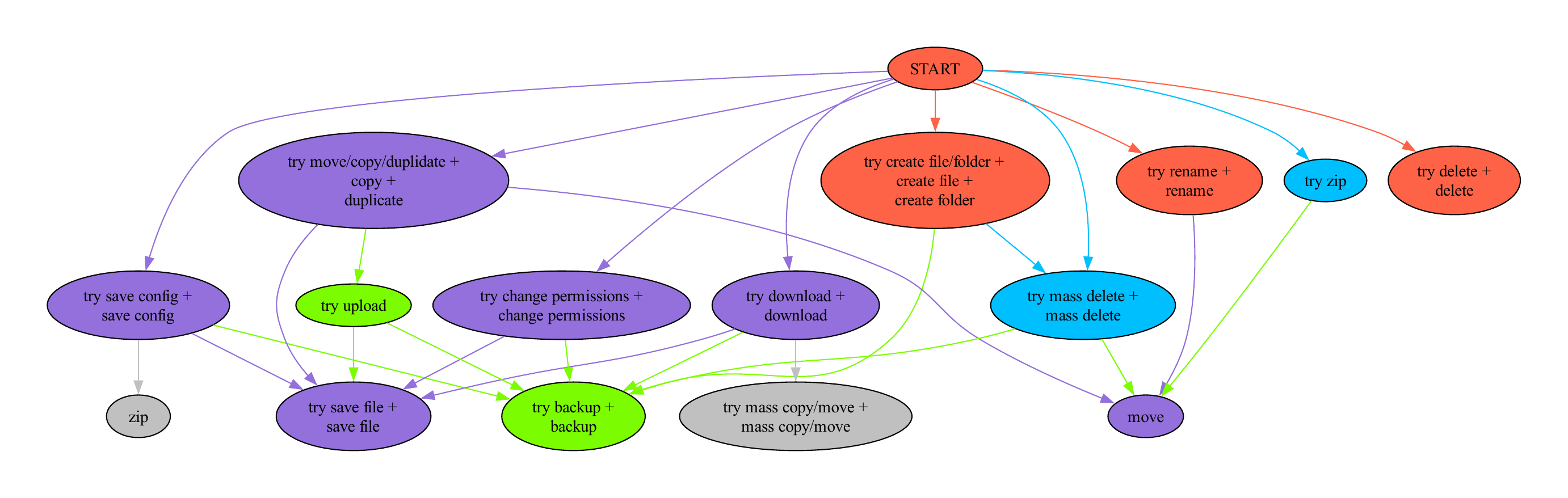}
	\caption{
		\tinyfilemanager state graph, using LoC that are manually defined as state-changing.
		These states are colored similarly to \Cref{fig:hotcrp_states}, except gray.
		Gray states are instead only found by \zap.}
	\label{fig:tinyfilemanager_states}
\end{figure}

In \Cref{fig:tinyfilemanager_states} we construct a state graph for \tinyfilemanager, based on our manual annotations of state-changing lines of code.
These manual annotations are listed in \Cref{tab:tinyfilemanager_annotations}.

\begin{table}
	\caption{
		Manual \tinyfilemanager annotations of state-changing LoC.
	}
	\label{tab:tinyfilemanager_annotations}
	\begin{center}
		\begin{tabular} {l | l}
			\toprule
			Annotation & LoC \\
			\midrule
START & \texttt{(index.php, 4)} \\
try login & \texttt{(index.php, 333)} \\
login & \texttt{(index.php, 337)} \\
search & \texttt{(index.php, 473)} \\
try save file & \texttt{(index.php, 480)} \\
save file & \texttt{(index.php, 501)} \\
try backup & \texttt{(index.php, 512)} \\
backup & \texttt{(index.php, 526)} \\
try save config & \texttt{(index.php, 538)} \\
save config & \texttt{(index.php, 573)} \\
change password & \texttt{(index.php, 580)} \\
try upload & \texttt{(index.php, 946)} \\
upload & \texttt{(index.php, 662)} \\
try delete & \texttt{(index.php, 676)} \\
delete & \texttt{(index.php, 685)} \\
try create file/folder & \texttt{(index.php, 699)} \\
create file & \texttt{(index.php, 710)} \\
create folder & \texttt{(index.php, 719)} \\
try move/copy/duplidate & \texttt{(index.php, 736)} \\
move & \texttt{(index.php, 761)} \\
copy & \texttt{(index.php, 769)} \\
duplicate & \texttt{(index.php, 793)} \\
try mass copy/move & \texttt{(index.php, 808)} \\
mass copy/move & \texttt{(index.php, 862)} \\
try rename & \texttt{(index.php, 877)} \\
rename & \texttt{(index.php, 896)} \\
try download & \texttt{(index.php, 910)} \\
download & \texttt{(index.php, 934)} \\
upload1 & \texttt{(index.php, 1024)} \\
upload2 & \texttt{(index.php, 1039)} \\
upload3 & \texttt{(index.php, 1062)} \\
try mass delete & \texttt{(index.php, 1091)} \\
mass delete & \texttt{(index.php, 1112)} \\
try zip & \texttt{(index.php, 1127)} \\
zip & \texttt{(index.php, 1176)} \\
try unzip & \texttt{(index.php, 1191)} \\
unzip & \texttt{(index.php, 1247)} \\
try change permissions & \texttt{(index.php, 1261)} \\
change permissions & \texttt{(index.php, 1309)} \\
			\bottomrule
		\end{tabular}
	\end{center}
\end{table}

\end{document}